\documentclass{aa}  

\usepackage{graphicx}
\usepackage{ulem}
\usepackage[mathscr]{euscript}
\usepackage{epsfig}
\usepackage{color}
\usepackage{txfonts}
\usepackage{lscape}
\usepackage{natbib}
\usepackage{soul}
\usepackage{longtable}
\usepackage{captcont}
\usepackage{pdfpages}
\usepackage{amsmath}
\usepackage{color}
\usepackage{soul, xcolor}
\bibpunct{(}{)}{;}{a}{}{,} 
\setcitestyle{citesep={,}}

\newcommand{\am}[2]{$#1'\,\hspace{-1.7mm}.\hspace{.0mm}#2$}

\newcommand{\HI}{H\textsc{i}}
\newcommand{\HIbf}{\mbox{H\hspace{0.155 em}{\footnotesize \bf I}}}

\newcommand{\MHI}{$M_{\rm HI}$}

\newcommand{\Mz}{$M_{\rm z}$}
\newcommand{\Mg}{$M_{\rm g}$}
\newcommand{\Mstar}{$M_{\star}$}

\newcommand{\Msun}{$M_\odot$}
\newcommand{\kms}{\mbox{km\,s$^{-1}$}}
\newcommand{\nan}{Nan\c{c}ay}
\def\approxlt{\lower.2em\hbox{$\buildrel < \over \sim$}}
\def\approxgt{\lower.2em\hbox{$\buildrel > \over \sim$}}

\newcommand{\FHI}{\mbox{$F_{\rm HI}$}}
\newcommand{\Jykms}{\mbox{Jy~km~s$^{-1}$}}
\newcommand{\kmsMpc}{\mbox{km~s$^{-1}$~Mpc$^{-1}$}}

\newcommand{\Lsun}{\mbox{$L_{\odot}$}}

\newcommand{\fMHIMstar}{\mbox{$\frac{{M}_{\rm HI}}{{M}_{\star}}$}}

\newcommand{\VHI}{\mbox{$V_{\rm HI}$}}
\newcommand{\Vopt}{\mbox{$V_{\rm opt}$}}

\newcommand{\Wfifty}{\mbox{$W_{\mathrm 50}$}}
\newcommand{\Wtwenty}{\mbox{$W_{\mathrm 20}$}}
\newcommand{\Lr}{\mbox{$L_{\rm r}$}}

\newcommand{\lamda}{\lambda}

\newcommand{\fML}{\mbox{$\frac{M_{\rm HI}/M_{\odot}}{ L_{\rm r}/L_{\odot}}$}}
\newcommand{\fMHIMsun}{\mbox{$\frac{{M}_{\rm HI}}{{M}_{\odot}}$}}
\newcommand{\fsSFR}{\mbox{$\frac{\rm sSFR}{\rm yr^{-1}}$}}
\newcommand{\fLrLsun}{\mbox{$\frac{L_{\rm r}}{L_{\odot}}$}}
\defcitealias{butcher2016}{Paper II}
\defcitealias{vandriel2016}{Paper I}
\defcitealias{butcher2018}{Paper III}
\begin{document}
\setstcolor{red}
    
\offprints{Z. Butcher}

\title{The NIBLES bivariate luminosity--\HI\ mass distribution function revised using Arecibo follow-up observations\thanks{Tables A1 and A2, along with the \HI\ line spectra are available in electronic form at the CDS via anonymous ftp to cdsarc.u-strasbg.fr (130.79.128.5) or via http://cdsweb.u-strasbg.fr/cgi-bin/qcat?J/A+A/}}
\author{Z. Butcher\inst{1}
        \and
        W. van Driel\inst{2,3} 
        \and
        S. Schneider\inst{1}
    }
    
\institute{University of Massachusetts, Astronomy Program, 536 LGRC, Amherst, MA 01003, U.S.A. 
 \email{zbutcher@umass.edu}
\and
GEPI, Observatoire de Paris, PSL Universit\'e, CNRS, 5 place Jules Janssen, 92190 Meudon, France  
\and
USN, Station de Radioastronomie de \nan, Observatoire de Paris, CNRS/INSU USR 704, Universit\'e d'Orl\'eans OSUC, route de Souesmes, 18330 \nan, France  
}
    
\date{Received  ; Accepted  }

\abstract{We present a modified optical luminosity--\HI\ mass bivariate luminosity function (BLF) based on \HI\ line observations from the \nan\ Interstellar Baryons Legacy Extragalactic Survey (NIBLES), including data from our new, four times more sensitive follow-up \HI\ line observations obtained with the Arecibo radio telescope. The follow-up observations were designed to probe the underlying \HI\ mass distribution of the NIBLES galaxies that were undetected or marginally detected in \HI\ at the \nan\ Radio Telescope. Our total follow-up sample consists of 234 galaxies, and it spans the entire luminosity and color range of the parent NIBLES sample of 2600 nearby (900 $< cz <$ 12,000 \kms) SDSS galaxies.  We incorporated the follow-up data into the bivariate analysis by scaling the NIBLES undetected fraction by an Arecibo-only distribution. We find the resulting increase in low \HI\ mass-to-light ratio densities to be about 10\% for the bins $-1.0 \le $ log(\fML) $ \le -0.5$, which produces an increased \HI\ mass function (HIMF) low mass slope of $\alpha = -1.14 \pm 0.07$, being slightly shallower than the values of $-1.35 \pm 0.05$ obtained by recent blind \HI\ surveys. Applying the same correction to the optically corrected bivariate luminosity function from our previous paper produces a larger density increase of about 0.5 to 1 dex in the lowest \HI\ mass-to-light ratio bins for a given luminosity while having a minimal effect on the resulting HIMF low mass slope, which still agrees with blind survey HIMFs.  This indicates that while low \HI-mass-to-light ratio galaxies do not contribute much to the one-dimensional HIMF, their inclusion has a significant impact on the densities in the two-dimensional distribution.}
    
    \keywords{
        Galaxies: statistics --
        Galaxies: general --
        Galaxies: formation --
        Galaxies: dwarf --
        Radio lines: galaxies   
    }
    
    \authorrunning{Z. Butcher}    
    \titlerunning{The revised NIBLES bivariate luminosity--\HI\ mass distribution function}
    \maketitle

\section{Introduction}

Understanding the large-scale baryonic mass distribution of the Universe has been one of the longest-standing quests in modern astronomy. The present-day baryonic mass distribution provides insight into many different aspects of the evolution of the Universe as it is the current endpoint for all evolutionary models.  This included but is not limited to, the evolution of dark matter haloes, how galaxies process gas within galaxies, stellar evolution and heavy element production, and galaxy mergers and interactions.  The spatial distribution of dark matter haloes is traced by galaxies whose sizes and morphologies provide constraints on the physical processes undergone by the baryons.

The most common and widely used quantifier of the galaxy population has historically been the optical luminosity function (LF), which describes the volume density of galaxies as a function of luminosity \citep[see, e.g.,][and \citealp{lan2016}]{loveday1992, marzke1994, norberg2002, blanton2003, dorta09, geller2012, roberts2014, loveday2015, parsa2016}.  Since the LF is a fundamental tracer of the galaxy population, it is commonly used as a test for semi-analytic galaxy formation models \citep[see, e.g.,][and \citealp{cooray2005}]{white1991, katz1992, kauffmann1993, cole1994, somerville1999, cole2000, pearce2001, benson2003}.

Similar to the optical LF, the \HI\ mass function (HIMF) has also been used to trace the galaxy population, although to a somewhat lesser extent \cite[see, e.g.,][and \citealp{hoppmann2015}]{zwaan97, schneider1998, zwaan2003, kovac2005, springob05, martin2010}.  While both of these population tracers are used as tests of semi-analytic models, they both only trace a single dimension of the galaxy population.

The optical luminosity--\HI\ mass bivariate luminosity function (BLF) describes the volume density of galaxies as a function of both their luminosity and \HI\ mass. When integrated over all luminosities in each \HI\ mass bin, it becomes the HIMF, and when integrated over all \HI\ masses in each luminosity bin, it becomes the optical LF.  The advantage of analyzing the galaxy population in more than one dimension simultaneously is that it provides details on the distribution of one variable as a function of the other, allowing for finer tuning of formation models.

Previously, \cite{lemonias2013} presented a bivariate distribution as a function of stellar and \HI\ masses.  However it was essentially stellar mass folded into a one-dimensional HIMF and it covered a far smaller mass range than our sample.
 	
The \nan\ Interstellar Baryons Legacy Extragalactic Survey (NIBLES), being an optically selected survey, is able to probe \HI\ masses to lower levels than typical blind \HI\ surveys due to increased on-source integration times.  This provides the advantage of being able to probe \HI\ mass distributions as a function of optical luminosity much better than blind \HI\ surveys, which have no optical selection criteria.  Conversely, blind \HI\ surveys tend to contain more \HI\ rich, low surface brightness (LSB) galaxies, which are often difficult to detect or completely absent from optical surveys.  The practical implications of these selection criteria are that HIMFs from optically selected sources tend to have flatter slopes than their blind survey counterparts \citep[see, e.g.,][]{rao1993}.  However, for the purposes of constructing a BLF, the \HI\ distribution as a function of luminosity is the critical component and is more readily obtained from optical selection criteria.

In \citet[][\citetalias{butcher2018}]{butcher2018}, we present the first optical luminosity--\HI\ mass BLF and HIMF of the NIBLES sample of 2600 galaxies selected from the Sloan Digital Sky Survey (SDSS; see, e.g., \citealt{york2000}) within the local universe (900 $\leq cz \leq $ 12,000 \kms). The galaxies were chosen with the goal of obtaining nearly equal numbers of galaxies in each absolute magnitude bin, so that the optical luminosity function was sampled more uniformly than in magnitude-limited studies or blind \HI\ surveys. The project was based on uniform \HI-line observations carried out with the 100-m class \nan\ Radio Telescope (NRT).

The \nan\ observations had a detection rate of 63\%, that is, 1497 out of the 2364 target sources for which usable \HI\ spectra were obtained and which were not clearly confused by another galaxy within the telescope beam \citep[][\citetalias{vandriel2016}]{vandriel2016}.  
The undetected galaxies are mainly high-luminosity, gas-poor red objects, and low-luminosity, predominantly blue, galaxies (see also \citetalias{butcher2018}). Although the low \HI\ mass sources within a given luminosity bin have very little impact on the overall distribution of the one-dimensional HIMF \citepalias[see][]{butcher2018}, they do alter the two-dimensional BLF distribution based on both optical luminosity and \HI\ mass.

We identified trends in the \HI-mass-to-luminosity (gas-to-light hereafter) ratios, \fML, that are consistent over the entire luminosity range of NIBLES, from which we extrapolated volume densities for galaxies with very low luminosities that lie outside the NIBLES selection criteria due to the SDSS magnitudes being generally unreliable for $r$-band apparent mag $> 17.77$ (a consequence of the SDSS selection criteria chosen to correspond to their desired target density of 90 objects per square degree, see \citealt{loveday2002}). The resulting extrapolated distributions produce a BLF from which we can derive an HIMF that is consistent with blind \HI\ survey HIMFs. 

To better understand the properties of the galaxies undetected at \nan\ and assess their impact on the BLF, we obtained four times higher sensitivity follow-up observations with the 305 m Arecibo radio telescope.  
In two earlier Arecibo follow-up campaigns \citep[see][\citetalias{butcher2016}, for details]{butcher2016} a total of 92 NIBLES galaxies were observed, from a random subsample and a subsample of nearby ($cz < 4000$ \kms) blue galaxies with a color $u-z<2$ mag. We present here results of a third Arecibo observing campaign of 151 objects (including some repeats from the earlier campaigns) designed to provide a fairly uniform sampling of, collectively, 234 NIBLES galaxies that were either not detected or marginally detected at \nan.

In Section \ref{sec:sample_selection} we describe the observed galaxy sample; in Section \ref{sec:observations} we describe the observation strategy and data reduction procedure; in Section \ref{sec:method_and_results} we review the methodology used to derive BLFs and HIMFs; in Section \ref{sec:discussion} we compare results including the Arecibo sample to the results of the \nan\ only sample; and in Section \ref{sec:conclusions} we present our conclusions. 
Solar luminosities are given in the SDSS $r$-band, derived from Christopher Willmer's calibrations \footnote{\url{http://mips.as.arizona.edu/~cnaw/sun_2006.html}}.

\section{Sample selection}
\label{sec:sample_selection}

The larger collecting area of the Arecibo radio telescope allows us to reach lower \HI\ mass limits for undetected NIBLES galaxies within the Arecibo declination range. Not all undetected galaxies could be observed due to telescope time scheduling constraints, so we aimed to expand the set of our earlier Arecibo observations to provide a collective subsample covering the entire NIBLES luminosity range in a fairly uniform manner, with less color bias than our earlier campaigns. In total, about half the undetected galaxies within  Arecibo's declination range were observed.  

The third, and final, Arecibo \HI\ line follow-up observation sample described here consists of 151 NIBLES galaxies that were observed from July 2016 to February 2017 during a total of 58.5 hours of telescope time. 132 of these galaxies were classified as \nan\ nondetections and 22 were marginal detections.  Of these, 9 were previously observed during the first two follow-up campaigns but resulted in nondetections.  They were observed in this campaign with longer integration times, resulting in \HI\ line detections of two sources, 1132 and 1983 (See \citetalias{butcher2016} for details on the first two campaigns).  Since the second campaign targeted blue galaxies ($u-z<2$ mag) these were excluded in this final campaign, which consists mostly of galaxies with redder colors and at greater distances than those in the previous campaigns. We refer to the combined set of observations from all three campaigns as the Arecibo sample hereafter.

\begin{figure}  
    \centering
    \includegraphics[width=9cm]{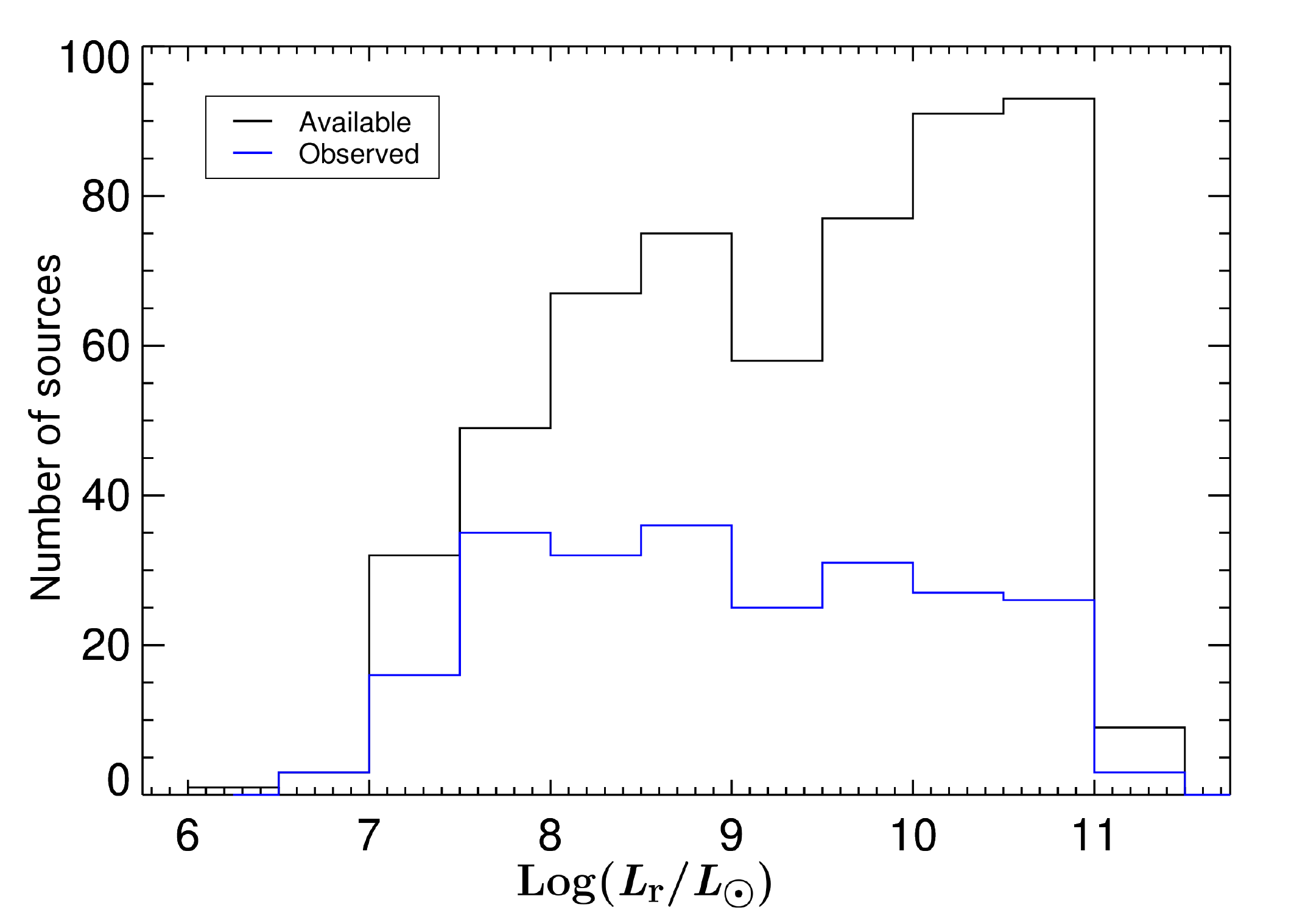}
    \caption{Luminosity distribution (log(\Lr/\Lsun)) of NIBLES galaxies within the Arecibo declination range that were undetected or marginally detected at \nan\ (black), along with the sources that were observed in the Arecibo follow-up campaigns (blue).}
    \label{fig:Lr_dist}
\end{figure}

\begin{figure} 
\centering
\includegraphics[width=9cm]{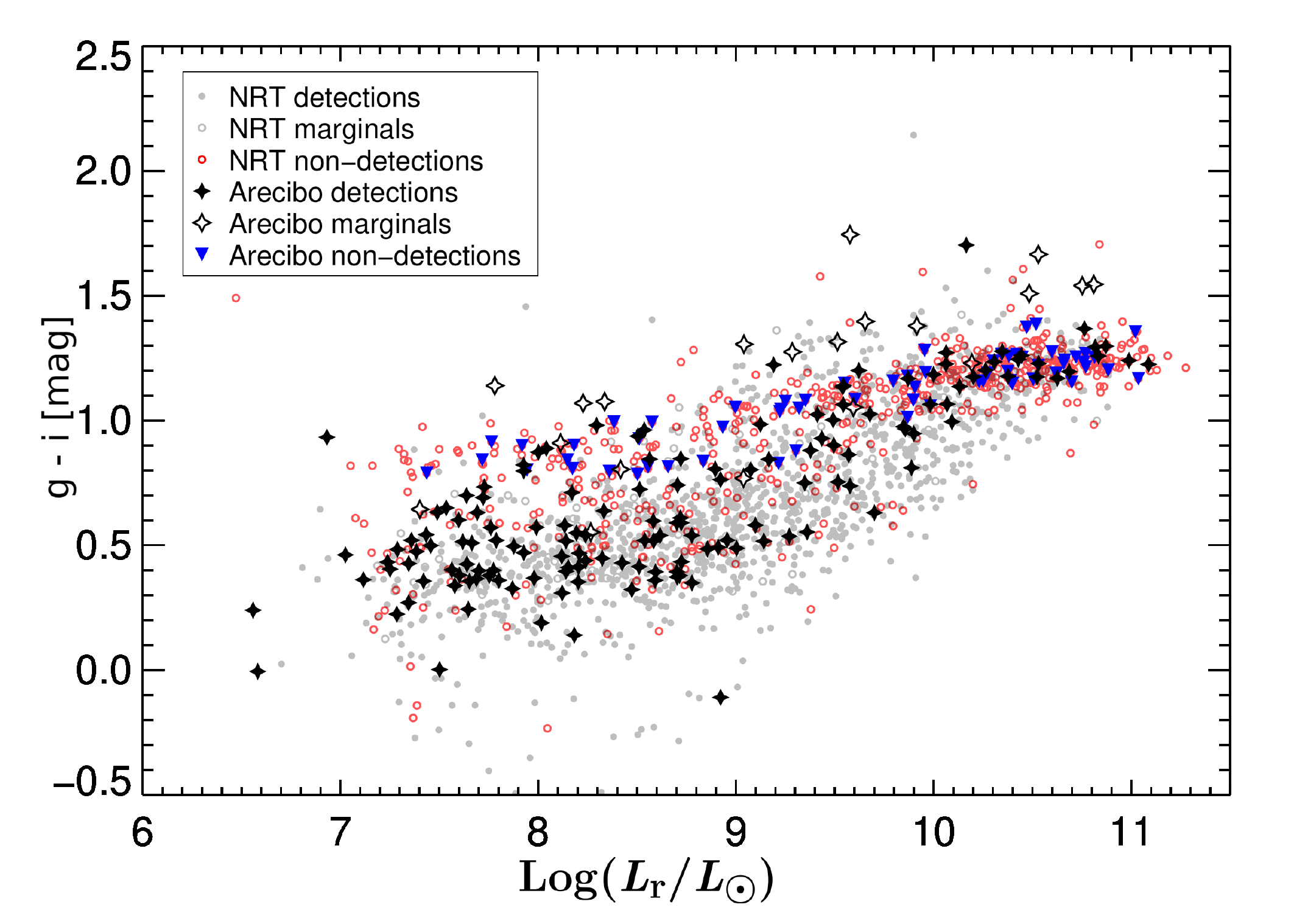}
\caption{\label{g_i_Lr_BLF} Integrated $g-i$ color, in mag, as a function of absolute $r$-band luminosity log(\Lr/\Lsun), both corrected for Galactic extinction following \cite{schlegel1998}. \nan\ detections, marginals, and nondetections are represented by gray dots, open gray circles, and open red circles respectively. Arecibo detections, marginals, and nondetections are respectively represented by black solid stars, open stars, and blue downward triangles.}
\label{fig:g_i_Lr_BLF}
\end{figure}

The luminosity and color distributions of the Arecibo sample are shown in Figures \ref{fig:Lr_dist} and \ref{fig:g_i_Lr_BLF} respectively. The latter shows the lack of a color bias, but the former shows that the Arecibo sample has an almost constant number ($\sim$25) of objects per luminosity bin, whereas the ensemble of possible Arecibo targets shows a steady increase with luminosity, from about 30 at log(\Lr/\Lsun) $\sim$ 7 to about 90 at log(\Lr/\Lsun) $\sim$ 11. The decreased percentage of observed sources at higher luminosities does not detrimentally impact our analysis since our requirement is to have higher sensitivity follow-up observations covering the entire luminosity range that are a fair representation of the overall NIBLES sample.  Since our observed sources were only limited by telescope availability, and they fully sample the color distribution shown in Figure \ref{fig:g_i_Lr_BLF}, the only impact on our analysis is the magnitude of uncertainty due to sampling errors.

\section{Observations and results}
\label{sec:observations}

The observing strategy of the third campaign is identical to that of our previous two at the Arecibo radio telescope (see \citetalias{butcher2016}). We used the L-wide receiver and the Wideband Arecibo Pulsar Processor (WAPP) autocorrelator backend with two polarizations, a bandpass of 50 MHz (10,600 \kms) and 4096 frequency channels separated by 2.6 \kms. The receiver half power beam width is \am{3}{5} and the system temperature ranges from 28 to 32 K. Data were taken in standard 5/5 minute integration $ON/OFF$ position switching mode. All galaxies were first observed for one 5/5 minute cycle, and the weak and nondetections were then repeated depending on available telescope time. 

Data were reduced using a combination of Phil Perillat's IDL routines and Robert Minchin's CORMEASURE routine from the Arecibo Observatory. All spectra were Hanning smoothed to a velocity resolution of 18.7 \kms\ to match the 18 \kms\ resolution of the NRT spectra as closely as possible. 

Throughout this paper, all recession velocities given are heliocentric, all \HI-line related parameters are according to the conventional optical definition: $V$ = c($\lamda$ -- $\lamda_0$)/$\lamda_0$, and a Hubble constant of $H_0 = 70$ \kmsMpc\ is used.  Galaxy properties and explanations of derived quantities are listed in Appendix \ref{sec:appendix}.

Classification of galaxies into \HI\ detected, nondetected and marginal categories was determined by the three authors, making independent judgments about how each galaxy should be classified based on visual inspection of the \HI\ spectra, with the final adjudication based on a majority consensus. Visual inspection was used rather than a uniform cut in signal-to-noise ratio because the optical redshift of each galaxy was known beforehand, allowing better recognition of source signals.

Color images along with \HI\ line spectra of all 151 galaxies in the third campaign are shown in Appendix \ref{sec:appendix}, in Figures \ref{fig:DET1} to \ref{fig:ND1}, along with Tables \ref{tab:A3060_DET}, \ref{tab:A3060_MAR} and \ref{tab:A3060_ND} listing data for clear detections, marginal detections and nondetections, respectively. 

Six sources had unreliable SDSS photometry or were confused by another galaxy within the Arecibo telescope beam (Nos. 0492, 0538, 0748, 0987, 2356 and 2483); these are flagged accordingly in Col. 1 of Tables \ref{tab:A3060_DET}, \ref{tab:A3060_MAR} and \ref{tab:A3060_ND}. Excluding these six leaves 228 sources to be used for further analysis. Of these, 64\% were clearly detected, and 8\% marginally detected. 
The detection fraction depends on color: all of the Arecibo marginal and nondetections are on the redder end of the color distribution ($g-i>0.8$), Figure \ref{fig:g_i_Lr_BLF}. 

\section{\Lr--\MHI\ bivariate luminosity function and \HIbf\ mass function}
\label{sec:method_and_results}

The basic methodology we use to derive an \Lr--\MHI\ BLF is the same as in \citetalias{butcher2018}. For each luminosity bin, we first count the number of galaxies in each \HI\ mass bin to determine the \HI\ mass distribution, and then scale the results to obtain the volume density required to match the known luminosity function from \cite{dorta09}. This gives us the two-dimensional volume density as a function of both \HI\ mass and luminosity.

In \citetalias{butcher2018} we derived an uncorrected bivariate \Lr--\HI\ mass distribution using \nan\ detections only, and a corrected distribution based on observed trends in the \MHI/\Lr\ distribution as a function of luminosity, which we used to extrapolate the distribution down to luminosities well below those of the NIBLES sample galaxies. For the new analysis presented here, which includes our Arecibo follow-up data, we consider only the uncorrected distribution and we treat marginal Arecibo detections as nondetections. 
 
 \begin{figure} 
\centering
\includegraphics[width=8.5cm]{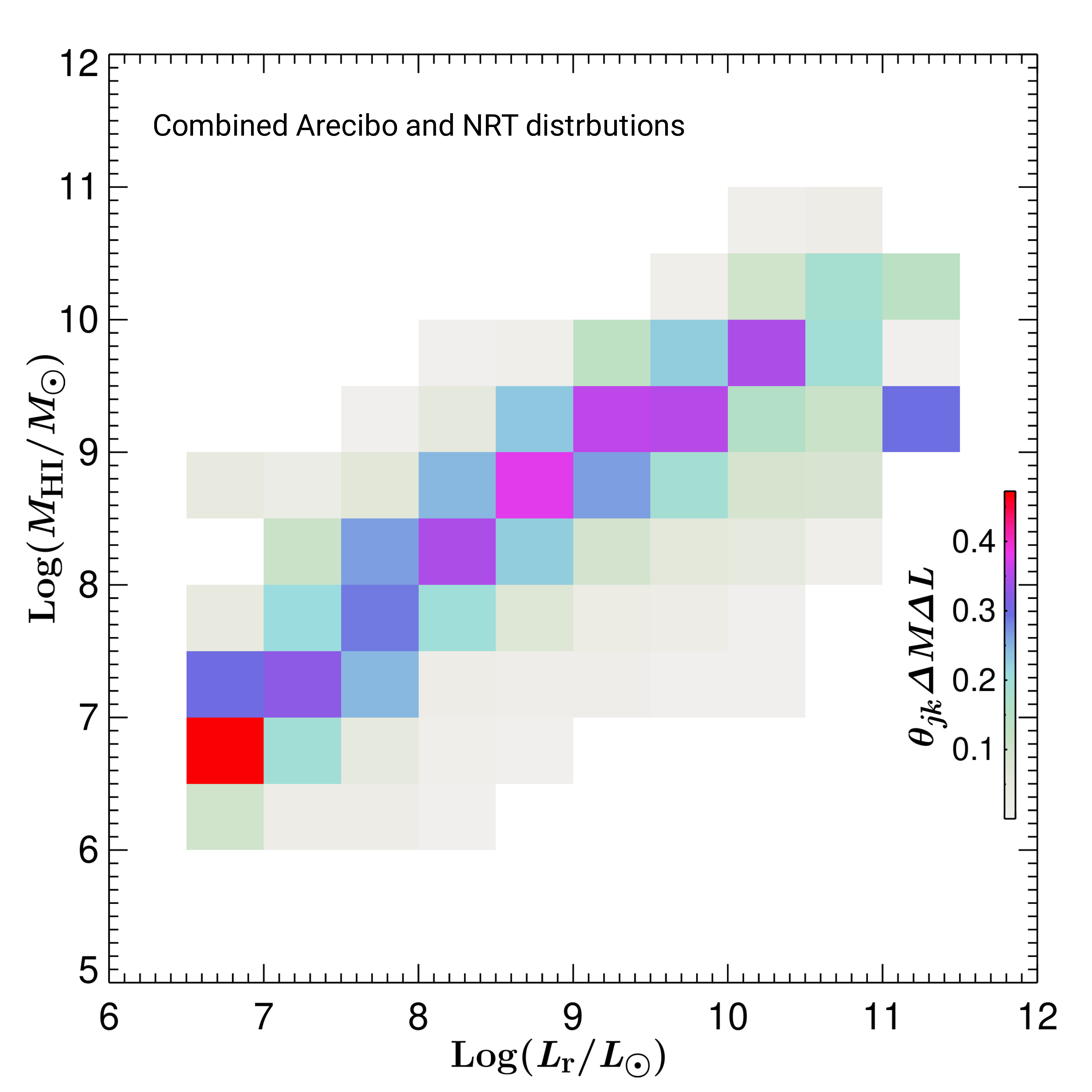}
\caption{\label{MHILrdens_combined} Two dimensional $r$-band luminosity--\HI\ mass distribution of the NIBLES sample, derived by combining the scaled two dimensional Arecibo distribution with the \nan\ distribution. The color scale indicates the fraction of galaxies that have a particular log(\MHI/\Msun) for a given log(\Lr/\Lsun), see the legend.}
\label{fig:MHILrdens_combined}
\end{figure}

 \begin{figure} 
\centering
\includegraphics[width=8.5cm]{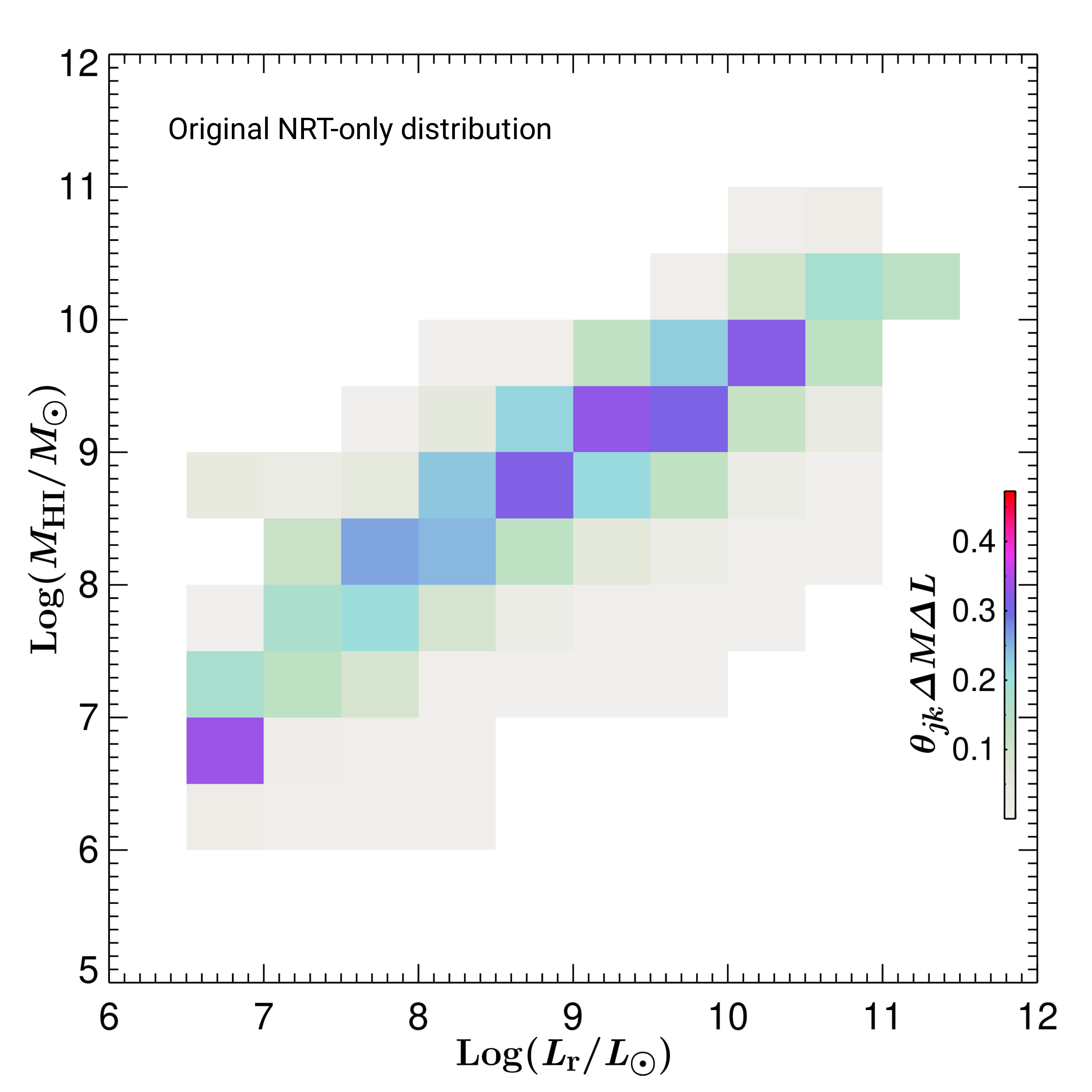}
\caption{\label{MHILrdens_orig} Two dimensional $r$-band luminosity--\HI\ mass distribution of the NIBLES sample derived using the \nan\ observations only -- from \citetalias{butcher2018}, with the same color scale shown in Figure \ref{fig:MHILrdens_combined}.}
\label{fig:MHILrdens_orig}
\end{figure}

 \begin{figure} 
\centering
\includegraphics[width=8.5cm]{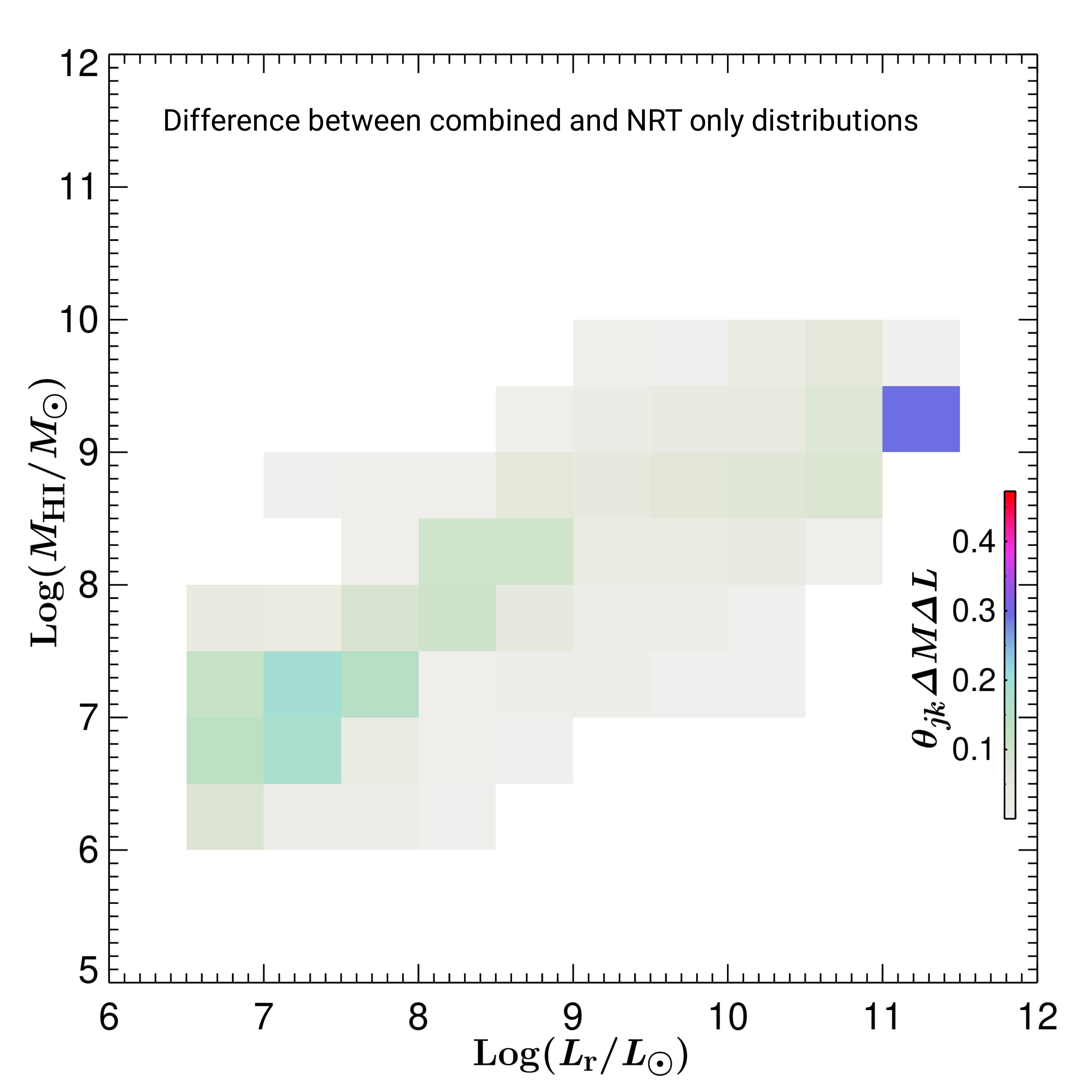}
\caption{\label{MHILrdens_diff} Difference between Figures \ref{fig:MHILrdens_combined} and \ref{MHILrdens_orig}, showing the impact of adding the Arecibo follow-up observations data to the original \nan\ data.}
\label{fig:MHILrdens_diff}
\end{figure}

The Arecibo sample provides a fairly uniform subset of undetected NIBLES galaxies within the Arecibo telescope's more limited declination range.  Therefore, we can assume that the \Lr--\HI\ mass distribution present in the Arecibo sample is a fair representation of all the \nan\ undetected galaxies. 

We characterize the Arecibo sample's \Lr--\HI\ mass distribution in the same manner as for the \nan\ sample. To do this, we follow the same general method outlined in \citetalias{butcher2018}, at first using only the Arecibo data. Specifically, we generate a distribution:
\begin{equation}
\theta_{Ajk} \Delta M \Delta L = n_{Ajk}/N_k,
\end{equation}
\noindent where $\theta_{Ajk}$ represents the distribution of galaxies in the $j$th \HI\ mass bin and $k$th luminosity bin of the Arecibo sample only, $n_{Ajk}$ is the total number of galaxies in the $jk$ \HI\ mass and luminosity bin, and $N_k$ is the total number of galaxies in luminosity bin $k$. 

To combine this distribution with the \nan\ sample distribution (see \citetalias{butcher2018}), we simply add the Arecibo distribution to the \nan\ distribution, scaling all $k$ luminosity bins of the Arecibo distribution by the fraction of undetected galaxies in the corresponding $k$ luminosity bin of the \nan\ distribution. 
The resulting combined distribution is shown in Figure \ref{fig:MHILrdens_combined} and 
the original \nan-only distribution in Figure \ref{MHILrdens_orig}, using the same color scale. The difference between the two distributions is shown in Figure \ref{MHILrdens_diff}.

As with the equivalent \nan\ detections-only distribution (Fig. \ref{fig:MHILrdens_orig}), the lowest \HI\ mass bin in Fig. \ref{fig:MHILrdens_combined} is populated exclusively by partial occupation numbers and therefore has a much lower density than the adjacent bins. 

The difference between the two distributions is that the Arecibo observations have increased the density of the BLF bins by approximately 10\% for bins corresponding to the colored regions of Figure \ref{MHILrdens_combined}. The lower luminosity bins, around log(\Lr/\Lsun) = 7.25, show an approximately 20\% increase in density and an almost 30\% increase for the bin log(\Lr/\Lsun) = 11.25, log(\MHI/\Msun) = 9.25.

The log(\Lr/\Lsun) $> 11$ galaxies likely require observations that are even more sensitive than ours at Arecibo to fully sample the \HI\ mass distribution within the 11.25 bin. The log(\Lr/\Lsun) = 11.25 bins in the BLF show that bin log(\MHI/\Msun) = 9.75 has a relatively lower density than the two adjacent \HI\ mass bins. This is due to insufficient sampling, with the 9.75 bin populated solely by fractional occupation numbers due to the uncertainties in the two adjacent bins. At log(\Lr/\Lsun) $> 11$, the \nan\ NIBLES sample contains only 15 galaxies, three of which were detected at \nan, all within the log(\MHI/\Msun) = 10.25 bin. Of the three Arecibo sample galaxies in the same luminosity bin, the sole detection was in the log(\MHI/\Msun) = 9.25 bin. Using the Arecibo sample as an estimator for the \nan\ distribution of undetected \HI\ galaxies leaves the log(\MHI/\Msun) = 9.75 bin populated by only partial occupation numbers based on \HI\ mass uncertainties in the adjacent bins as described in \citetalias{butcher2018}.

\begin{figure} 
\centering
\includegraphics[width=8.5cm]{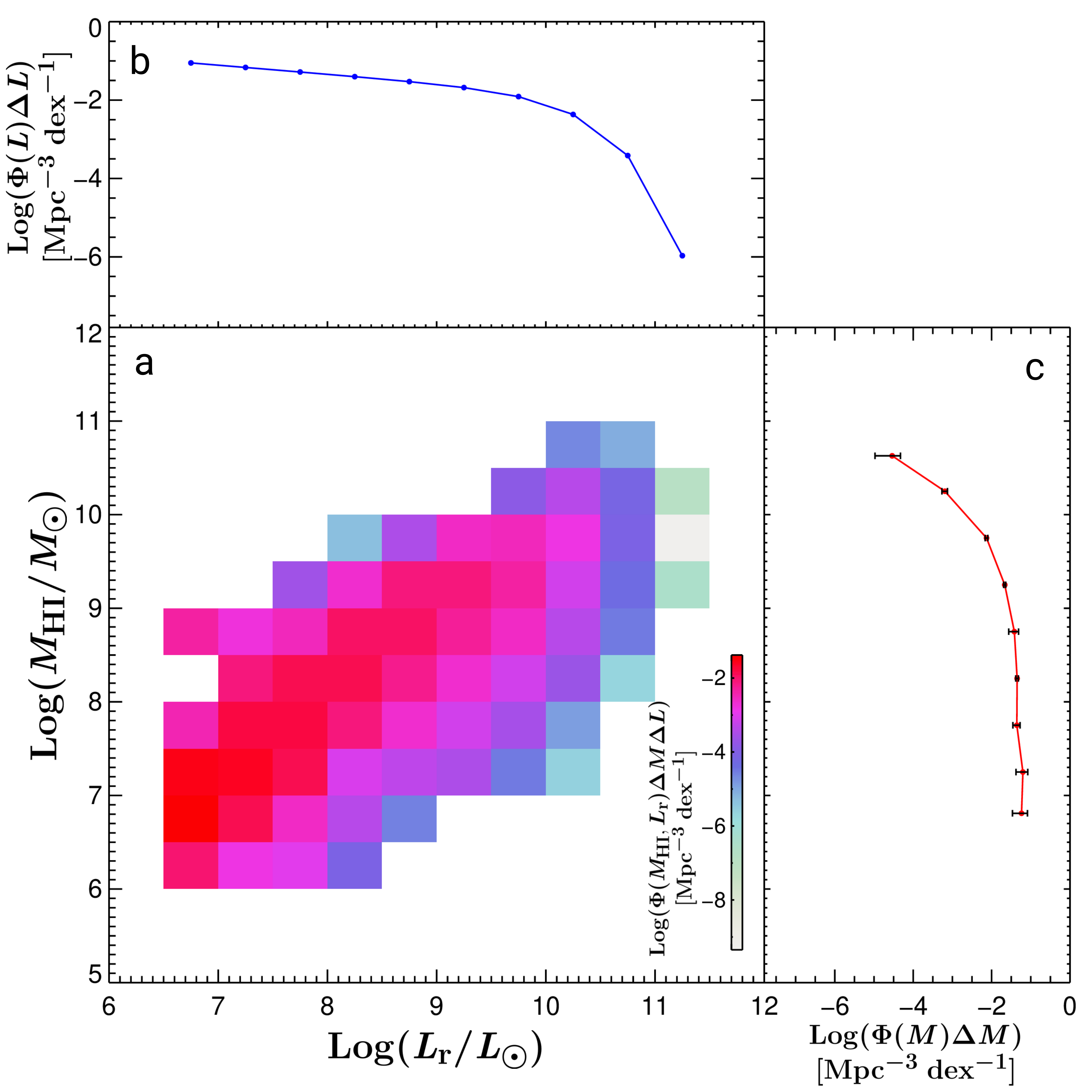}
\caption{\label{MHILrspacedens_combined} Main panel $a$: Two-dimensional bivariate \Lr--\HI\ mass distribution of the NIBLES sample galaxies, derived from a combination of the \nan\ and Arecibo sample distributions. The values represented by the colorbar (see the legend) are the volume densities in each 0.5 dex wide bin in both luminosity and \HI\ mass,
log($\Phi$(\MHI,\Lr)$\Delta M \Delta L$) in units of Mpc$^{-3}$ dex$^{-1}$, as a function of both \HI\ mass and $r$-band luminosity \Lr\ (in solar units). The upper panel $b$ shows the summation of the main panel over \HI\ mass, which reproduces the input Luminosity Function from \citet{dorta09}, log($\Phi(L)\Delta L$) in units of Mpc$^{-3}$ dex$^{-1}$, while the right panel $c$ shows the summation of the bivariate distribution over luminosity, that is, the \HI\ Mass Function, log($\Phi(M)\Delta M$) in units of Mpc$^{-3}$ dex$^{-1}$. See the text for further details on the log(\MHI/\Msun) = 6.25 and 10.75 bins in the HIMF.}
\label{fig:MHILrspacedens_combined}
\end{figure}

Scaling each luminosity bin from Figure \ref{fig:MHILrdens_combined} to the corresponding bin from the \citet{dorta09} luminosity function (LF) 
yields the two-dimensional $r$-band luminosity--\HI\ mass distribution function shown in panel $a$ of Figure \ref{fig:MHILrspacedens_combined}, and its corresponding HIMF (panel $c$).  Similarly to Figure 4 in \citetalias{butcher2018}, we did not plot the point for the lowest \HI\ mass bin (log(\MHI/\Msun) = 6.25) in the HIMF since it was populated with partial occupation numbers only, due to the relatively high fractional \HI\ mass uncertainty of these sources. The position for the highest \HI\ mass bin (log(\MHI/\Msun) = 10.75) in the HIMF is plotted at the mean value of the measured \HI\ masses contained within the bin rather than at its normal \MHI\ midpoint since this bin is not fully sampled (see Figure \ref{fig:MHILrspacedens_combined}).

Values for the mass distribution function from panel ($a$) in Figure \ref{fig:MHILrspacedens_combined} are listed in Table \ref{tab:BLF} as log($\Phi$(\MHI,\Lr)$\Delta M$ $\Delta L$) in units of Mpc$^{-3}$ dex$^{-1}$ (in solar units), together with their fractional uncertainties.

\begin{table*}  
\caption{\label{BLF} $r$-band luminosity--\HI\ mass distribution function for the uncorrected combined \nan\ and Arecibo distribution.} 
\begin{tabular}{lcccccccccc}
\hline
\!\!\!\!\!\! & \multicolumn{10}{c}{Log(\Lr/\Lsun) } \\
log(\fMHIMsun) \!\!\!\!\!\! &\!\!\!\!\!\! 6.75 &\!\!\!\!\!\! 7.25 &\!\!\!\!\!\! 7.75 &\!\!\!\!\!\! 8.25 &\!\!\!\!\!\! 8.75 &\!\!\!\!\!\! 9.25 &\!\!\!\!\!\! 9.75 &\!\!\!\!\!\! 10.25 &\!\!\!\!\!\! 10.75 &\!\!\!\!\!\! 11.25 \\
 \!\!\!\!\!\! &  \multicolumn{10}{c}{[volume density in Mpc$^{-3}$ dex$^{-1}$]} \\
\hline
\hline
10.75 \!\!\!\!&\!\!\!\!\!\!               &\!\!\!\!\!\!               &\!\!\!\!\!\!               &\!\!\!\!\!\!               &\!\!\!\!\!\!               &\!\!\!\!\!\!               &\!\!\!\!\!\!               &\!\!\!\!\!\! -4.70$\pm$0.42&\!\!\!\!\!\! -5.07$\pm$0.20 \\
10.25 \!\!\!\!&\!\!\!\!\!\!               &\!\!\!\!\!\!               &\!\!\!\!\!\!               &\!\!\!\!\!\!               &\!\!\!\!\!\!               &\!\!\!\!\!\!               &\!\!\!\!\!\! -3.86$\pm$0.23&\!\!\!\!\!\! -3.36$\pm$0.07&\!\!\!\!\!\! -4.15$\pm$0.06&\!\!\!\!\!\! -6.85$\pm$0.46 \\
9.75  \!\!\!\!&\!\!\!\!\!\!               &\!\!\!\!\!\!               &\!\!\!\!\!\!               &\!\!\!\!\!\! -5.31$\pm$2.27&\!\!\!\!\!\! -3.46$\pm$0.23&\!\!\!\!\!\! -2.57$\pm$0.07&\!\!\!\!\!\! -2.55$\pm$0.05&\!\!\!\!\!\! -2.83$\pm$0.04&\!\!\!\!\!\! -4.13$\pm$0.11&\!\!\!\!\!\! -9.53$\pm$14.39 \\
9.25  \!\!\!\!&\!\!\!\!\!\!               &\!\!\!\!\!\!               &\!\!\!\!\!\! -3.62$\pm$0.44&\!\!\!\!\!\! -2.69$\pm$0.11&\!\!\!\!\!\! -2.16$\pm$0.05&\!\!\!\!\!\! -2.13$\pm$0.04&\!\!\!\!\!\! -2.36$\pm$0.04&\!\!\!\!\!\! -3.17$\pm$0.08&\!\!\!\!\!\! -4.37$\pm$0.19&\!\!\!\!\!\! -6.50$\pm$0.35 \\
8.75  \!\!\!\!&\!\!\!\!\!\! -2.38$\pm$1.03&\!\!\!\!\!\! -2.78$\pm$0.29&\!\!\!\!\!\! -2.50$\pm$0.13&\!\!\!\!\!\! -2.01$\pm$0.05&\!\!\!\!\!\! -1.95$\pm$0.04&\!\!\!\!\!\! -2.26$\pm$0.05&\!\!\!\!\!\! -2.64$\pm$0.07&\!\!\!\!\!\! -3.42$\pm$0.15&\!\!\!\!\!\! -4.48$\pm$0.25 \\
8.25  \!\!\!\!&\!\!\!\!\!\!               &\!\!\!\!\!\! -2.13$\pm$0.12&\!\!\!\!\!\! -1.86$\pm$0.05&\!\!\!\!\!\! -1.86$\pm$0.05&\!\!\!\!\!\! -2.16$\pm$0.06&\!\!\!\!\!\! -2.71$\pm$0.10&\!\!\!\!\!\! -3.15$\pm$0.15&\!\!\!\!\!\! -3.70$\pm$0.22&\!\!\!\!\!\! -5.59$\pm$1.16 \\
7.75  \!\!\!\!&\!\!\!\!\!\! -2.36$\pm$0.85&\!\!\!\!\!\! -1.85$\pm$0.10&\!\!\!\!\!\! -1.82$\pm$0.06&\!\!\!\!\!\! -2.10$\pm$0.07&\!\!\!\!\!\! -2.67$\pm$0.13&\!\!\!\!\!\! -3.17$\pm$0.21&\!\!\!\!\!\! -3.57$\pm$0.29&\!\!\!\!\!\! -4.94$\pm$0.76&\!\!\!\!\!\!  \\
7.25  \!\!\!\!&\!\!\!\!\!\! -1.57$\pm$0.33&\!\!\!\!\!\! -1.65$\pm$0.10&\!\!\!\!\!\! -1.90$\pm$0.07&\!\!\!\!\!\! -3.06$\pm$0.27&\!\!\!\!\!\! -3.36$\pm$0.35&\!\!\!\!\!\! -3.45$\pm$0.31&\!\!\!\!\!\! -4.51$\pm$0.89&\!\!\!\!\!\! -5.75$\pm$2.73&\!\!\!\!\!\!  \\
6.75  \!\!\!\!&\!\!\!\!\!\! -1.39$\pm$0.24&\!\!\!\!\!\! -1.88$\pm$0.14&\!\!\!\!\!\! -2.59$\pm$0.19&\!\!\!\!\!\! -3.43$\pm$0.49&\!\!\!\!\!\! -4.61$\pm$1.48&\!\!\!\!\!\!               &\!\!\!\!\!\!               &\!\!\!\!\!\!               &\!\!\!\!\!\!     \\

\hline
\end{tabular}
\tablefoot{Volume densities are log($\Phi$(\MHI,\Lr)$\Delta M$ $\Delta L$) in Mpc$^{-3}$ dex$^{-1}$.  The listed uncertainties are fractional.}
\label{tab:BLF}
\end{table*}


\begin{figure}  
\centering
\includegraphics[width=8.5cm]{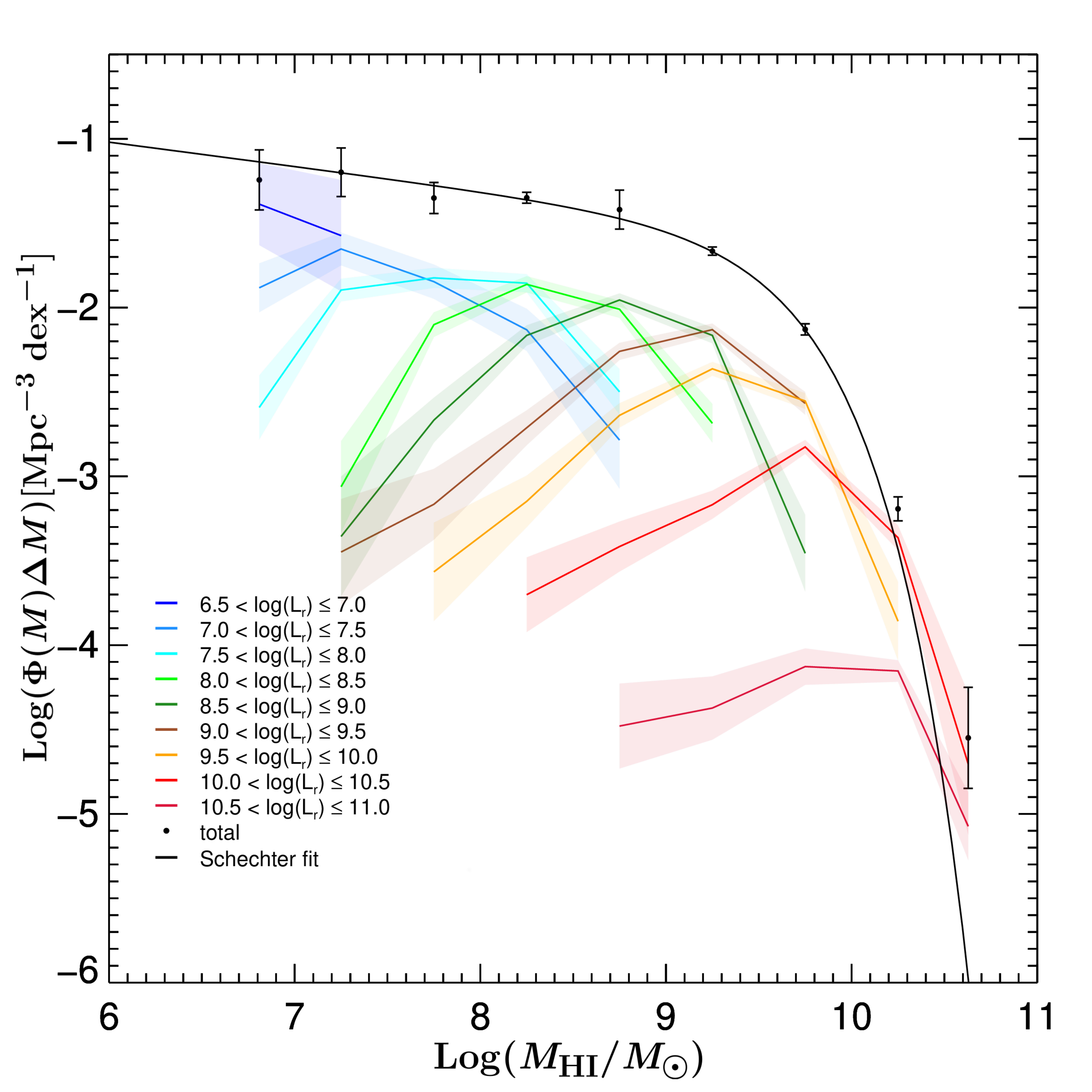}
\caption{\label{lumbins_r_combined} \HI\ Mass Function (data points and the Schechter function fit to them, in black), and contributions to the HIMF for individual luminosity bins, indicated by different colors (see the Legend). Volume densities in each 0.5 dex wide bin in \HI\ mass, log($\Phi$(M)$\Delta$M) in units of Mpc$^{-3}$ dex$^{-1}$, are shown as a function of \HI\ mass in each luminosity bin. For clarity, we omit data points with volume density values that are smaller than their uncertainties. The black points are the sum of the \HI\ masses in the corresponding luminosity bins. Uncertainties for each luminosity bin are shown as shaded regions around each mass function, with the total quadrature sum shown as error bars on the HIMF. The black line is the Schechter fit to the HIMF.}
\label{fig:lumbins_r_combined}
\end{figure}

In Figure \ref{fig:lumbins_r_combined} we show the contributions to the HIMF from panel $c$ of Figure \ref{fig:MHILrspacedens_combined} per luminosity bin. Due to the combination of Arecibo and \nan\ data, some of the \HI\ mass bins on the extremities of a particular luminosity bin have very large fractional uncertainties, being at or above unity (or above 0.434 on the logarithmic scale). We have omitted these points from the plot for viewing clarity, but their values are listed in Table \ref{tab:BLF}. Additionally, the sole data point in the log(\Lr/\Lsun) = 11.25 bin that is below the stated uncertainty threshold is also left off this plot due to its insignificant contribution to the HIMF.  Comparing this plot to Figure 5 in \citetalias{butcher2018} illustrates that the addition of the Arecibo data to the BLF results in an increased low-\HI-mass slope of the HIMF. A Schechter function \citep[see][]{schechter1976} fit to this HIMF yields the following parameters: \\

\noindent  $\Phi = 0.012 \pm 0.003$, log($M_{\star}$/\Msun) = $9.63 \pm 0.06$, $\alpha = -1.14 \pm 0.07$. \\

The corresponding HIMF based on NRT data only from \citetalias{butcher2018} yielded the following Schechter fit parameters: \\

\noindent  $\Phi = 0.013 \pm 0.002$, log($M_{\star}$/\Msun) = $9.61 \pm 0.06$, $\alpha = -1.04 \pm 0.07$. \\

The addition of the Arecibo data to the \nan\ distribution steepens the low-mass slope $\alpha$, from $-1.04$ to $-1.14$, which is expected given the high detection fraction for low luminosity sources in the Arecibo sample. The value of $\alpha$ for the uncorrected combined HIMF is still shallower than the low-mass slopes of blind \HI\ surveys ($-1.35 \pm 0.05$; \citealt{zwaan2005,haynes2011}) or the -1.26$\pm$0.04 of our optically corrected HIMF in Section 5.4 of \citetalias{butcher2018} (which used extrapolated gas-to-light distributions as a function of luminosity to construct distributions in luminosity bins not probed by the NIBLES sample --- see also Section \ref{sec:discussion}). 

\begin{figure}  
\centering
\includegraphics[width=8.5cm]{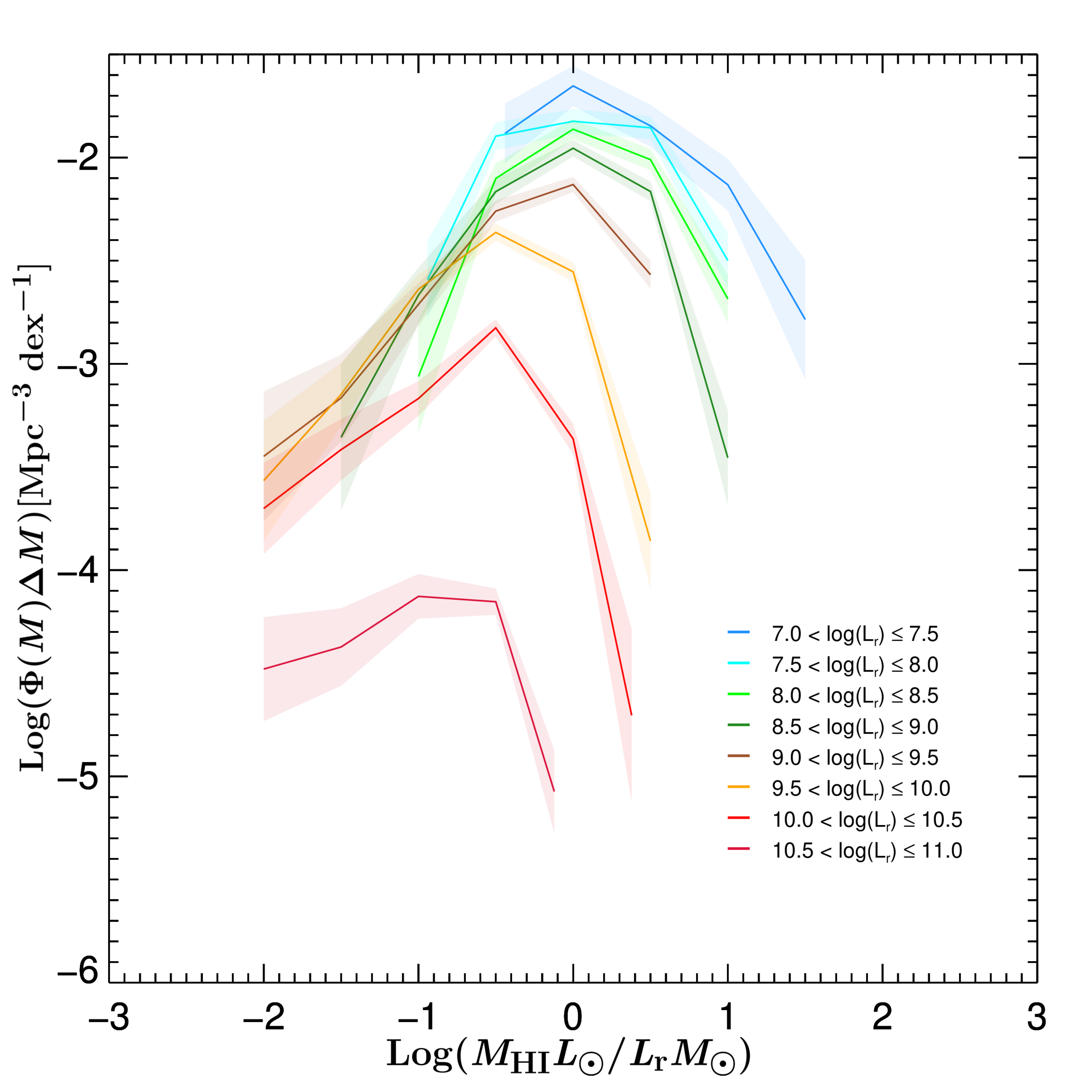}
\caption{\label{ML_ratio_combined} Volume densities in each 0.5 dex wide bin of \HI\ mass, log($\Phi(M)\Delta M$) in units of Mpc$^{-3}$ dex$^{-1}$, as a function of gas-to-light ratio, log(\fML) for the same luminosity bins as shown in Fig. \ref{fig:lumbins_r_combined}. }
\label{fig:ML_ratio_combined}
\end{figure}

In addition to the low-mass slope increase of the HIMF, within each luminosity bin, the density values corresponding to the lowest gas-to-light ratio bins have also increased. We show this effect more clearly in Figure \ref{fig:ML_ratio_combined}, where we can see that with the exception of the highest luminosity bin, the distributions of the low gas-to-light ratio bins (log(\fML) $\le -0.5$) all display similar shapes and slopes (this was also noted in Figure 6 of \citetalias{butcher2018}).  The additional detections from the Arecibo sample have increased the population density for the lowest gas-to-light ratios within each luminosity bin by similar amounts, with density increases of about 0.2 dex for log(\fML) values an order of magnitude below the peak, and about 0.4 dex density 1.5 orders of magnitude below the peak. 

The largest difference between Figure \ref{fig:ML_ratio_combined} and the equivalent Figure 6 of \citetalias{butcher2018} is that in Figure \ref{fig:ML_ratio_combined}, the maximum density values for log(\Lr/\Lsun) $\le 9.25$ all occur in the log(\fML) = 0 bin whereas in Figure 6 of \citetalias{butcher2018}, descending from log(\Lr/\Lsun) = 9.25 to 7.25 there is a progressive increase in the maximum density values from the log(\fML) = 0 bin to the 0.5 bin. The new density peaks are the result of adding the relatively lower gas-to-light ratio galaxies detected at Arecibo to the lower luminosity bins. On the other hand, in higher luminosity bins (log(\Lr/\Lsun) $\ge$ 9.75) the addition of low gas-to-light ratio objects did not alter the peak log(\MHI/Lr) bins.

The trend in gas-to-light ratios among higher luminosity galaxies is to be expected, and suggests that the more luminous galaxies are more evolved in the sense that they have converted progressively larger fractions of their gas into stars.  On the other hand, while the low-luminosity galaxies do not show a shift in their peak gas-to-light ratio, they do display successively larger numbers of galaxies with higher gas fractions at lower luminosities, whereas toward higher luminosities (log(\Lr/\Lsun) $> 9.5$) the higher gas-to-light ratio objects are gradually disappearing.

\section{Discussion}  
\label{sec:discussion}

In this section we examine how the Arecibo follow-up results (in particular the Arecibo nondetections) relate to the \nan\ results, and what we can infer from them about the properties of the NIBLES sample.

\begin{figure}  
\centering
\includegraphics[width=8.5cm]{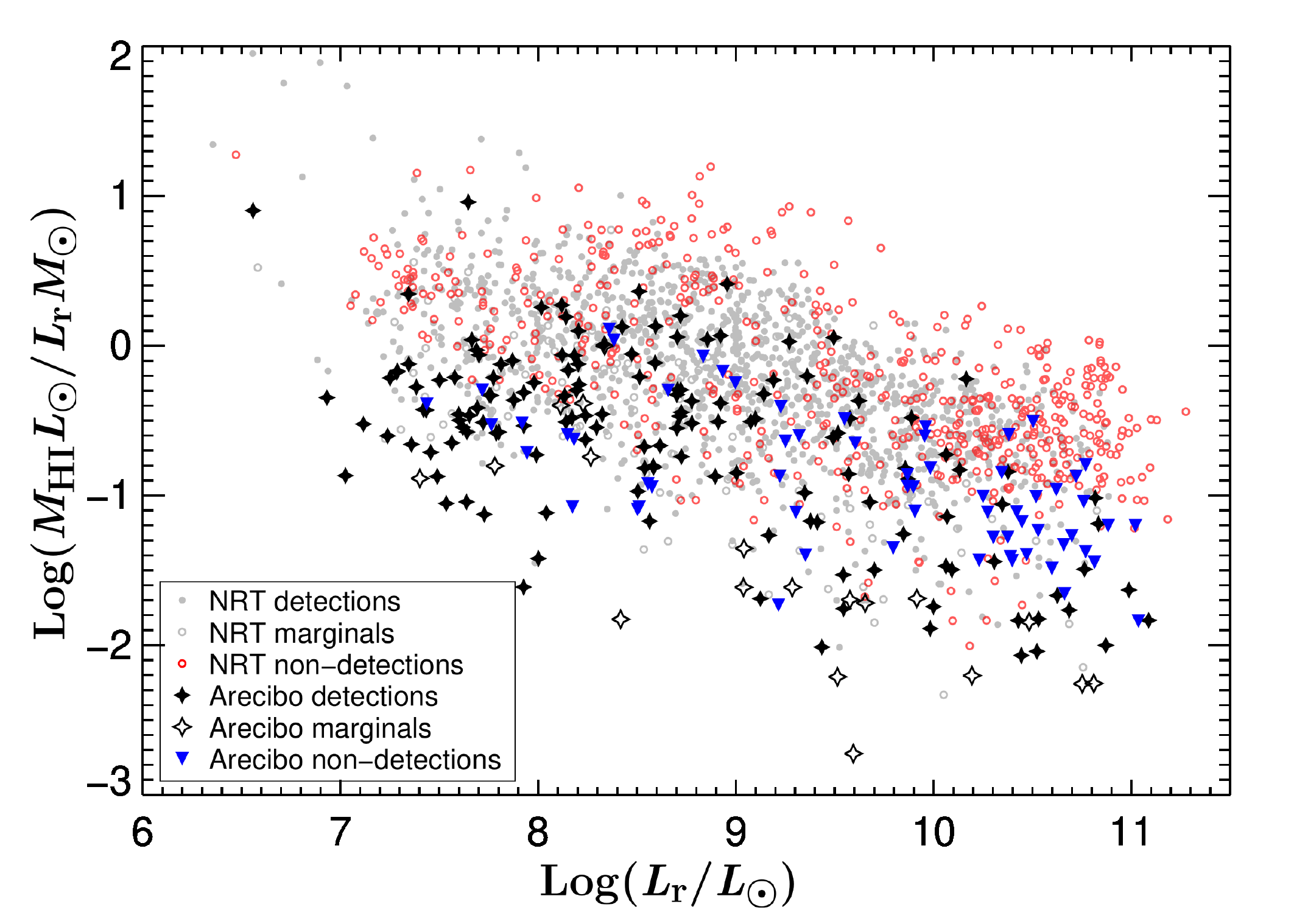}
\caption{\label{MHILr_Lr_data} Log(\fML), as a function of log(\Lr/\Lsun) for the Arecibo and \nan\ samples within the Arecibo sample's declination range.  Excluded are sources that were clearly confused or had unreliable photometry.  \nan\ detections, marginals, and nondetections are represented by gray dots, open gray circles, and open red circles respectively. Arecibo detections, marginals, and nondetections are respectively represented by black solid stars, open stars, and blue downward triangles. }
\label{fig:MHILr_Lr_data}
\end{figure}

\begin{figure}  
\centering
\includegraphics[width=8.5cm]{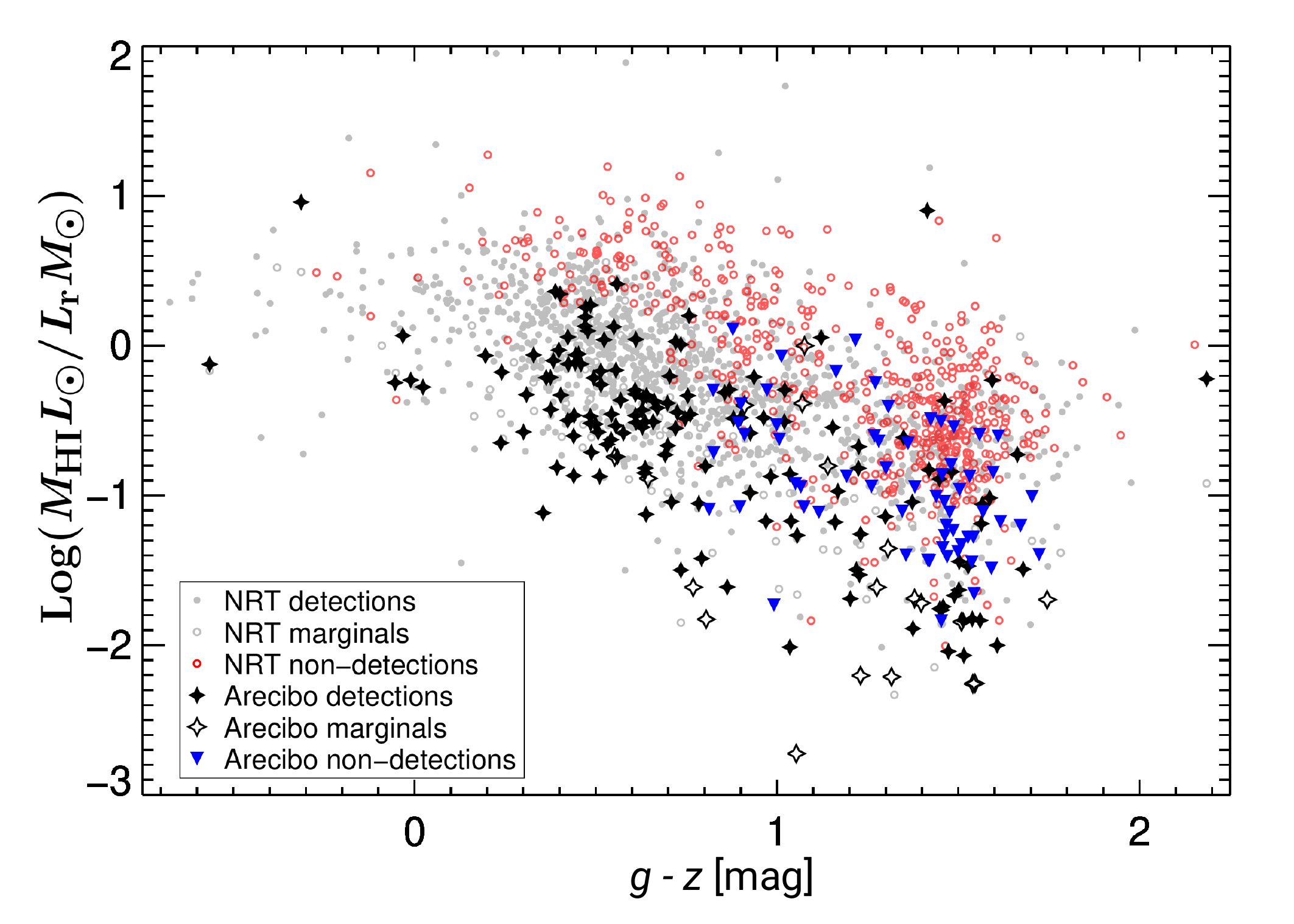}
\caption{\label{MHILr_color_data} Log(\fML) as a function of $g-z$ color for the Arecibo and \nan\ samples.  Excluded are sources that were clearly confused or had unreliable photometry. \nan\ detections, marginals, and nondetections are represented by gray dots, open gray circles, and open red circles respectively. Arecibo detections, marginals, and nondetections are respectively represented by black solid stars, open stars, and blue downward triangles. }
\label{fig:MHILr_color_data}
\end{figure}

Shown in Figures \ref{fig:MHILr_Lr_data} and \ref{fig:MHILr_color_data} are the gas-to-light ratios, log(\fML),  plotted as a function of $r$-band luminosity, \Lr, and $g-z$ color respectively. Excluded from the plots are the \nan\ data for sources of which Arecibo follow-up observations were obtained, and those that were clearly confused or had unreliable photometry.

It should be noted (see also \citetalias{vandriel2016}) that the estimated upper limits to the \HI\ masses of nondetections are quite conservative, as they are based on the largest observed \Wtwenty\ line widths for a given luminosity. For the most luminous sources, which are expected to have the broadest lines, the upper limits tend to be even higher than the NIBLES detections made with the same telescope at the same redshift (see Figure \ref{fig:MHILr_Lr_data}). 

The two Figures show that: 
(1) the gas-to-light ratios of the Arecibo nondetections lie below the mean for the \nan\ detections for each luminosity, that the mean difference between Arecibo and \nan\ nondetections is consistent with the four times higher Arecibo sensitivity, 
(2) about half of the Arecibo detections lie among the \nan\ detections, and the other half have on average about a ten times lower gas-to-light ratio, and 
(3) all galaxies blueward of $g-z = 0.75$ mag are detected in the Arecibo follow-up observations.

In \citetalias{butcher2018} we discussed the low-luminosity (log(\Lr/\Lsun) $< 7.25$) galaxies that were missing from the NIBLES sample due to insufficient sensitivity of optical surveys such as the SDSS and the effect this had on the resulting HIMF, based on \nan\ data only. Identifying trends as a function of luminosity allowed us to extrapolate the density and gas-to-light distribution values toward the low-luminosity bins with missing galaxies.
To analyze the effects of this change in detection fraction with luminosity we constructed two extrapolated BLFs, which we referred to as optically corrected and corrected, respectively. 
For the optically corrected BLF, we ignored the change in detection fraction as a function of luminosity and set the detection fraction of all log(\Lr/\Lsun) $< 7.75$ bins equal to that of the lowest luminosity well-sampled 7.75 bin.
For the corrected BLF, we included the change in detection fraction with luminosity and set the detection fraction for all log(\Lr/\Lsun) $< 9.25$ bins equal to that of the 9.25 bin because the fall-off in their detection fractions below this bin are consistent with decreases caused solely by distance and sensitivity effects \citepalias[see Appendix A in][]{butcher2018}. 
Our corrected HIMF agrees well with those derived from the HIPASS and ALFALFA blind surveys  \citet{zwaan2005,haynes2011}.

Here we expand the analysis to include the Arecibo sample's effect on the optically corrected bivariate distribution. We do not re-examine the corrected distribution, because it attempted to compensate for \HI\ undetected galaxies due to distance and sensitivity effects. Since our Arecibo data are four times more sensitive than the \nan\ data, we detect many of the \nan\ undetected galaxies that were the reason for the original correction.  The decrease in detection rate as a function of color in Figures \ref{fig:g_i_Lr_BLF} and \ref{fig:MHILr_color_data} corroborates this claim. Our 100\% detection rate for blue galaxies with $g-z < 0.75$ mag is not unexpected. Based on an analysis similar to that in Appendix A of \citetalias{butcher2018}, we would expect to detect $66 \pm 2$ of the 69 blue galaxies in our follow-up sample, that is, a detection rate of 96-99\%. This estimate is based on our minimum detectable integrated line flux at Arecibo and sampling errors from the standard deviation of the binomial distribution, but we do not have a uniform rms noise level due to our observing strategy where initially weak or nondetected sources received follow-up observations.

In contrast to the corrected distribution, the optically corrected distribution only attempts to correct for low-luminosity galaxies that were not included in the original NIBLES sample. In order to extrapolate the bivariate distribution across the full luminosity range (for which the reconstruction of a plausible HIMF was required) we must account for these missing low-luminosity galaxies in our new distribution utilizing the Arecibo follow-up data. 

\begin{figure}  
\centering
\includegraphics[width=8.5cm]{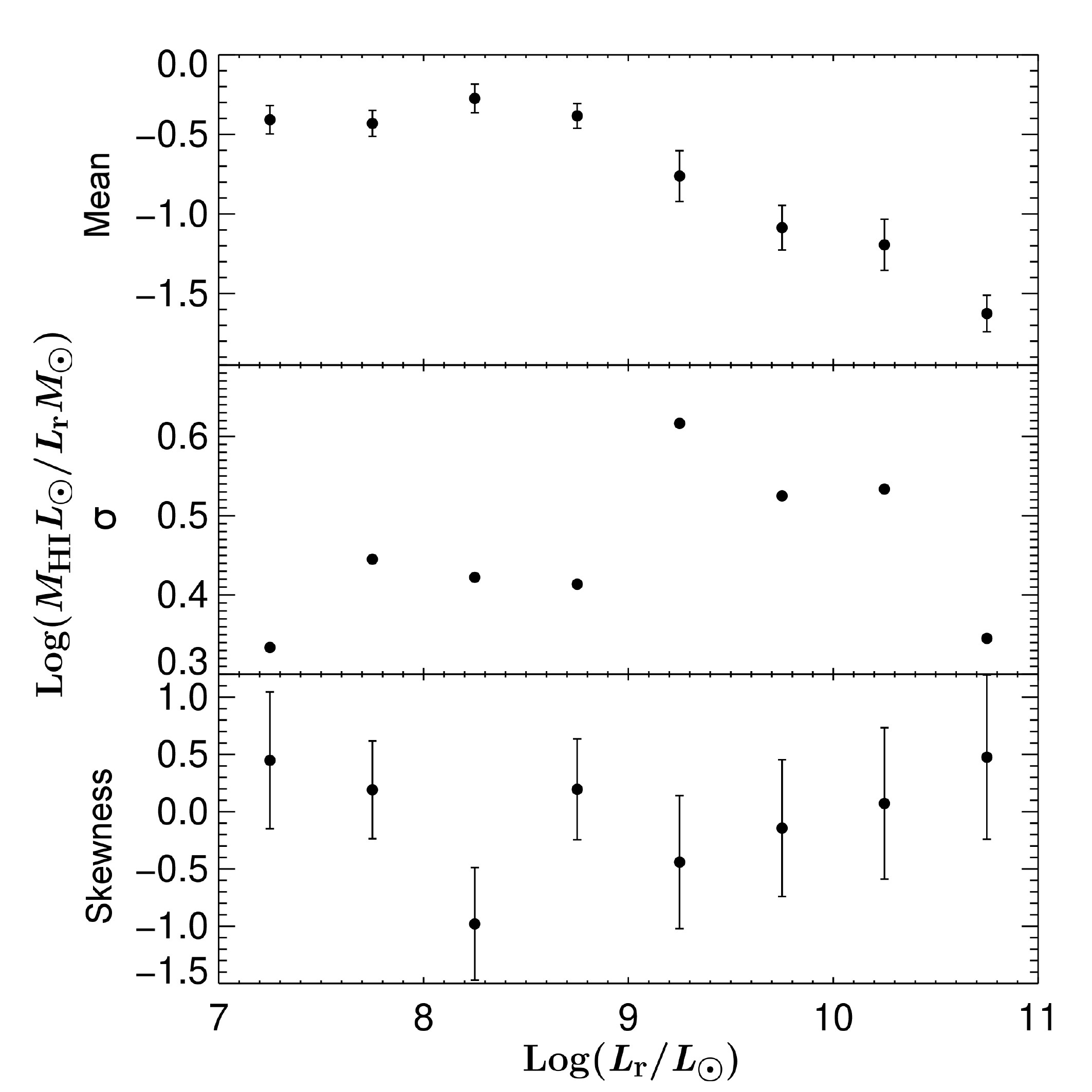}
\caption{\label{ML_ratio_combined_BLF_AO} Properties of the gas-to-light distributions of the Arecibo sample detections as a function of luminosity, in 0.5 dex wide bins of log(\Lr/\Lsun). Top panel: mean log(\fML) ratio, with uncertainties given by the standard error of the mean. Middle panel: standard deviation ($\sigma$) of the log(\fML) ratio. Lowest panel: Skewness of the log(\fML) ratio.}
\label{fig:ML_ratio_combined_BLF_AO}
\end{figure}

Unlike the \nan\ data (see Figure 8 in \citetalias{butcher2018}), the Arecibo data do not show any consistent trends with luminosity in the mean, standard deviation and skewness of the gas-to-light ratio over the entire luminosity range. The mean Arecibo gas-to-light ratios, standard deviations and skewness values in luminosity bins $7.25 \le $ log(\Lr/\Lsun) $\le 8.75$ all agree with one another within the uncertainties, and the gas-to-light ratio distributions can be accurately represented by a nonskewed Gaussian (see Figure \ref{fig:ML_ratio_combined_BLF_AO}). 

When correcting the \nan\ distributions, we extrapolated gas-to-light ratios for log(\Lr/\Lsun) $\le 7.25$ due to poor detection statistics in those bins. However, for the Arecibo follow-up data, we only extrapolate trends for luminosity bins log(\Lr/\Lsun) $\le 6.75$ due to the log(\Lr/\Lsun) = 7.25 bin now having a detection rate of 94\% in a sample size of 35. 

We first construct extrapolated gas-to-light distributions in low-luminosity bins down to log(\Lr/\Lsun) = 5.25 following the same general procedure outlined in Section 5.4 of \citetalias{butcher2018}, but using a nonskewed Gaussian such that within each luminosity bin we have: 

\begin{equation}
\mathfrak{R}_j = \int_{\Delta M/2L}^{-\Delta M/2L} \phi \\
\textrm{and} \\
\mathfrak{F}_k = \int_{min\ M/L}^{max\ M/L} \phi
\end{equation}
\noindent where $\mathfrak{R_j}$ is the $j$th gas-to-light ratio bin, $\phi$ represents the Gaussian function, and $\mathfrak{F_k}$ is the detection fraction in the $k$th luminosity bin. $\Delta M/L$ is the log(\fML) bin size.
    
We then combine the resulting distribution with the \nan-based distribution using the same procedure described in Section \ref{sec:method_and_results}. 

\begin{center}
\begin{table*}  
\caption{\label{BLF_corrected_combined} $r$-band luminosity--\HI\ mass distribution function for the corrected combined \nan\ and Arecibo distribution.}
\begin{tabular}{lccccccc}
\hline
\!\!\!\!\!\!& \multicolumn{7}{c}{Log(\Lr/\Lsun) } \\
log(\fMHIMsun)  \!\!\!\!\!\!& 5.25 \!\!\!\!& 5.75 \!\!\!\!& 6.25 \!\!\!\!& 6.75 \!\!\!\!& 7.25 \!\!\!\!& 7.75 \!\!\!\!& 8.25 \\
  \!\!\!\!\!\!& \multicolumn{7}{c}{[volume density in Mpc$^{-3}$ dex$^{-1}$]} \\
\hline
\hline
9.75\!\!\!\!\!\!&\!\!\!\!&\!\!\!\!&\!\!\!\!&\!\!\!\!&\!\!\!\!&\!\!\!\!&-5.31$\pm$2.28 \\
9.25\!\!\!\!\!\!&\!\!\!\!&\!\!\!\!&\!\!\!\!&\!\!\!\!&\!\!\!\!&-3.62$\pm$0.43\!\!\!\!&-2.68$\pm$0.11 \\
8.75\!\!\!\!\!\!&\!\!\!\!&\!\!\!\!&\!\!\!\!&\!\!\!\!&-3.08$\pm$0.35\!\!\!\!&-2.50$\pm$0.11\!\!\!\!&-2.01$\pm$0.04 \\
8.25\!\!\!\!\!\!&\!\!\!\!&\!\!\!\!&-3.35$\pm$0.51\!\!\!\!&-2.44$\pm$0.20\!\!\!\!&-2.00$\pm$0.04\!\!\!\!&-1.86$\pm$0.05\!\!\!\!&-1.86$\pm$0.06 \\
7.75\!\!\!\!\!\!&-3.70$\pm$0.69\!\!\!\!&-2.57$\pm$0.31\!\!\!\!&-1.95$\pm$0.08\!\!\!\!&-1.73$\pm$0.03\!\!\!\!&-1.76$\pm$0.04\!\!\!\!&-1.82$\pm$0.07\!\!\!\!&-2.10$\pm$0.10 \\
7.25\!\!\!\!\!\!&-1.97$\pm$0.16\!\!\!\!&-1.59$\pm$0.03\!\!\!\!&-1.57$\pm$0.04\!\!\!\!&-1.74$\pm$0.06\!\!\!\!&-1.73$\pm$0.13\!\!\!\!&-1.90$\pm$0.10\!\!\!\!&-3.07$\pm$0.17 \\
6.75\!\!\!\!\!\!&-1.36$\pm$0.04\!\!\!\!&-1.49$\pm$0.05\!\!\!\!&-1.75$\pm$0.10\!\!\!\!&-1.71$\pm$0.25\!\!\!\!&-1.91$\pm$0.18\!\!\!\!&-2.59$\pm$0.16\!\!\!\!&-3.43$\pm$0.22 \\
6.25\!\!\!\!\!\!&-1.48$\pm$0.07\!\!\!\!&-1.78$\pm$0.17\!\!\!\!&-1.62$\pm$0.29\!\!\!\!&-1.70$\pm$0.31\!\!\!\!&-2.84$\pm$0.34\!\!\!\!&-2.97$\pm$0.19\!\!\!\!&-4.08$\pm$0.26 \\
5.75\!\!\!\!\!\!&-1.82$\pm$0.26\!\!\!\!&-1.52$\pm$0.31\!\!\!\!&-1.56$\pm$0.31\!\!\!\!&-2.34$\pm$0.27\!\!\!\!&\!\!\!\!&\!\!\!\!& \\
5.25\!\!\!\!\!\!&-1.40$\pm$0.32\!\!\!\!&-1.42$\pm$0.30\!\!\!\!&-2.20$\pm$0.27\!\!\!\!&-3.75$\pm$0.25\!\!\!\!&\!\!\!\!&\!\!\!\!& \\
\hline
\end{tabular}
\tablefoot{Volume densities are log($\Phi$(\MHI,\Lr)$\Delta M$ $\Delta L$) in Mpc$^{-3}$ dex$^{-1}$.
The listed uncertainties are fractional.}
\label{tab:BLF_corrected_combined}	
\end{table*}
\end{center}

\begin{figure}  
\centering
\includegraphics[width=8.5cm]{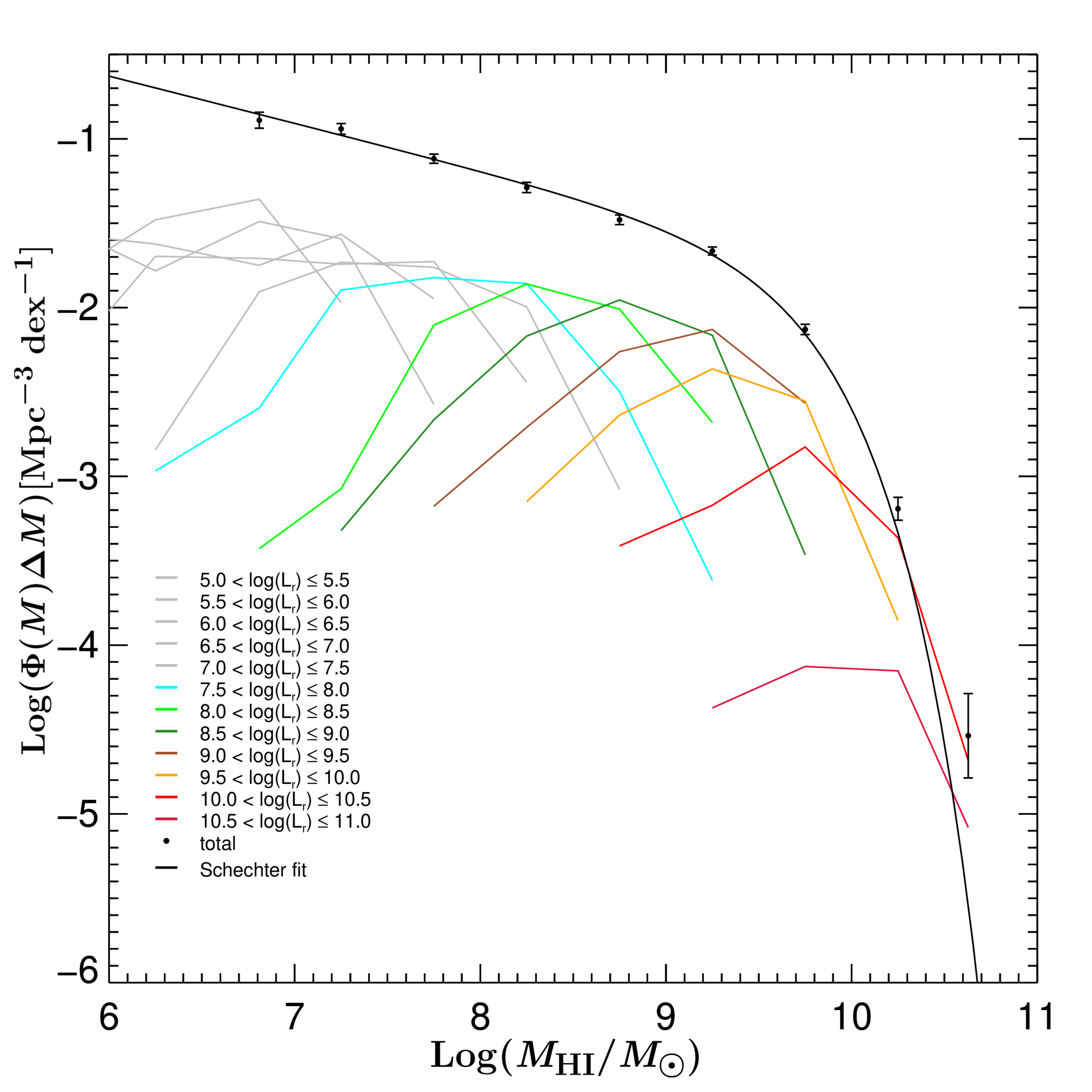}
\caption{\label{lumbins_r_fake_nanao} Reconstructed \HI\ Mass Function based on the combined optically corrected \nan\ and Arecibo distributions (data points and the Schechter function fit to them, in black), and contributions to the HIMF for individual luminosity bins, indicated by different colors (see the Legend). Volume densities are shown in each 0.5 dex wide bin in \HI\ mass,  log($\Phi(M)\Delta M$), in units of Mpc$^{-3}$ dex$^{-1}$. We recreated a plot similar to Figure \ref{fig:lumbins_r_combined} with artificial, extrapolated \HI\ mass distributions for luminosity bins below log(\Lr/\Lsun) = 7.75 shown in gray. For viewing clarity we omit points with uncertainties larger than the density values and do not show the uncertainty regions on each luminosity bin. Uncertainties for the corrected bins are given in Table \ref{tab:BLF_corrected_combined}.}
\label{fig:lumbins_r_fake_nanao}
\end{figure}

The resulting reconstructed HIMF is shown in Figure \ref{fig:lumbins_r_fake_nanao}. As with the combined \nan/Arecibo distribution shown in Figure \ref{fig:lumbins_r_combined}, we also omit the points on the extremities of the luminosity bins due to their large uncertainties and for viewing clarity. We present the values with their associated fractional uncertainties for luminosity bins log(\Lr/\Lsun) $\le 8.25$ in Table \ref{tab:BLF_corrected_combined}. 
The \HI\ mass distributions for the log(\Lr/\Lsun) $\le 7.75$ bins are more flattened than the corresponding bins in Figure 9 from \citetalias{butcher2018}, showing a density increase of $\sim$0.5 - 1 dex for the lowest \HI\ mass bins within each luminosity bin; due to the increased density of low log(\fML) ratio distributions. The Schechter fit parameters for the reconstructed HIMF are: \\

\noindent $\Phi = 0.009 \pm 0.002$, log($M_{\star}$/\Msun) = $9.70 \pm 0.06$, $\alpha = -1.28 \pm 0.03$. \\

This result agrees very well with the optically corrected HIMF from \citetalias{butcher2018}: \\

$\Phi = 0.0085 \pm 0.0015$, log($M_{\star}$/\Msun) = $9.72 \pm 0.06$, $\alpha = -1.26 \pm 0.04$ \\

\noindent and with the HIMF from \citet{zwaan2003}: \\

$\Phi = 0.0086$, log($M_{\star}$/\Msun) = $9.79$, $\alpha = -1.30.$ \\

While our ``corrected'' distribution (accounting for drop-offs in detection statistics due to distance and sensitivity effects at \nan) from \citetalias{butcher2018} produced an HIMF that agreed well with the HIMF from \cite{zwaan2005}, our corrected distribution presented here (using Arecibo data in lieu of the sensitivity correction) agrees better with the \cite{zwaan2003} results, showing that our increased detection statistics in the low luminosity bins did not drastically alter the low mass slope of the HIMF in comparison with the ``optically corrected'' distribution from \citetalias{butcher2018}.  

The similarity between the reconstructed HIMFs both with and without the Arecibo data illustrates the point that the HIMF is defined by the relatively higher gas-to-light ratio sources within each luminosity bin.  Therefore, the details of the gas-to-light distributions as a function of optical luminosity are lost when examining only a single dimension such as the LF or HIMF.  The only way to fully quantify the volume density of galaxies in terms of both \HI\ mass and optical luminosity, is to use a two-dimensional distribution.  This point should be of particular interest to galaxy evolution modelers as it illustrates a missing dimension in attempts to fit models to observed distributions. For example, a multitude of semi-analytic model studies have been conducted \citep[see, e.g.,][]{white1991, katz1992, kauffmann1993, cole1994, cole2000, somerville1999, pearce2001, benson2003, cooray2005} that use either a one-dimensional LF or HIMF as a basis for comparisons with observations. Comparing models with a two-dimensional distribution could help fine-tune various aspects of evolutionary models that is not possible when examining only a single dimension. 
 
\section{Conclusions}
\label{sec:conclusions}

We presented data from the third and final Arecibo \HI\ follow-up campaign of \nan\ nondetections of NIBLES galaxies, combining this data with our previous two Arecibo campaigns to create a random subsample spanning the range in luminosity and $g-i$ color of the NIBLES sample. We used this follow-up data to create a two dimensional bivariate luminosity--\HI\ mass distribution that we then scaled by the \HI\ undetected fractions of the \nan\ uncorrected and optically corrected distributions. 

Combining the resulting scaled Arecibo distribution with our original \nan\ distributions resulted in a net density increase of about 10\% at the low gas-to-light ratio bins ($-1.0<$log(\fML)$<-0.5$) within each luminosity bin. The effect on the uncorrected bivariate distribution resulted in a HIMF with a steeper low mass slope, but one that is just shallow of recent blind survey HIMFs (with the slope disagreeing just outside the uncertainty range). 

However, the bins with the lowest luminosities for the optically corrected BLF saw density increases of $\sim$0.5 to 1 dex for the highest gas-to-light ratio bins while producing a minimal change to the resulting HIMF low mass slope. This result illustrates the point that there may be significant deviations in the volume density for galaxies on the outskirts of the gas-to-light ratio distributions, as a function of environment or other factors, which we are currently unable to probe with a one-dimensional distribution.

Our results confirm that low gas-to-light ratio galaxies contribute relatively little to the overall \HI\ volume density of the universe and that optically selected samples tend to lack adequate numbers of high gas-to-light ratio galaxies from which to construct a realistic HIMF. However, the low gas-to-light ratio galaxies provide valuable insight into the gas-to-light distributions of the overall galaxy population and will aid our understanding of evolutionary processes within these galaxies. 

\begin{acknowledgements}
The Arecibo Observatory is a facility of the National Science Foundation operated under cooperative agreement (AST-1744119) by the University of Central Florida in alliance with Universidad Ana G. Méndez (UAGM) and Yang Enterprises (YEI), Inc.

The \nan\ Radio Telescope is operated as part of the Paris Observatory, in association with the Centre National de la Recherche Scientifique (CNRS) and partially supported by the R\'egion Centre in France. 

Funding for SDSS-III has been provided by the Alfred P. Sloan Foundation, the Participating Institutions, the National Science Foundation, and the U.S. Department of Energy Office of Science. The SDSS-III web site is http://www.sdss3.org/.

SDSS-III is managed by the Astrophysical Research Consortium for the Participating Institutions of the SDSS-III Collaboration including the University of Arizona, the Brazilian Participation Group, Brookhaven National Laboratory, Carnegie Mellon University, University of Florida, the French Participation Group, the German Participation Group, Harvard University, the Instituto de Astrofisica de Canarias, the Michigan State/Notre Dame/JINA Participation Group, Johns Hopkins University, Lawrence Berkeley National Laboratory, Max Planck Institute for Astrophysics, Max Planck Institute for Extraterrestrial Physics, New Mexico State University, New York University, Ohio State University, Pennsylvania State University, University of Portsmouth, Princeton University, the Spanish Participation Group, University of Tokyo, University of Utah, Vanderbilt University, University of Virginia, University of Washington, and Yale University. 

We would like to thank the anonymous referee for their helpful comments and constructive feedback to improve this article.

\end{acknowledgements}    
    
\bibstyle{aa}
\bibliographystyle{aa}
    
\bibliography{NIBLES_thesis}

\appendix
\section{Arecibo \HI\ line data}  
\label{sec:appendix}

\begin{figure} 
\centering
\includegraphics[width=15cm]{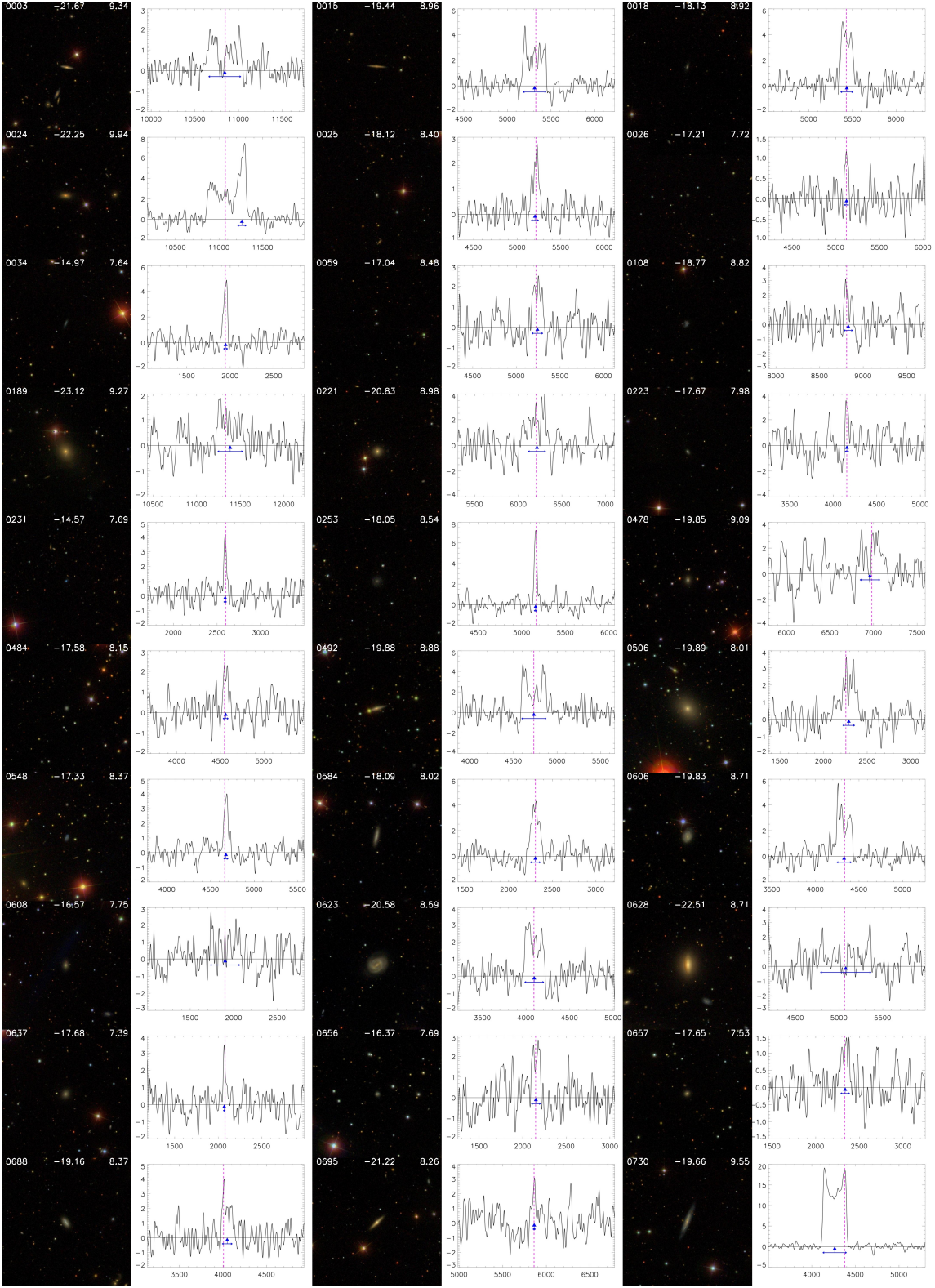}
\caption{\label{DET1} Color ($g$,$r$ and $i$ band composite) images from the SDSS alongside 21-cm \HI\ line spectra of galaxies clearly detected at Arecibo. The size of each image is 2\arcmin $\times$ 2\arcmin with the NIBLES source number indicated in the upper left corner, absolute z-band magnitude, \Mz, in the top center and log(\MHI/\Msun) in the top right corner of each image. The vertical axis in the spectra is flux density in mJy, the horizontal axis is heliocentric recession velocity ($cz$) in \kms. The SDSS recession velocity is denoted by a vertical dashed magenta line, the mean \HI\ velocity by the blue triangle, and the \Wfifty\ line width by the horizontal blue arrow bar. Velocity resolution is 18.7 \kms. Confused galaxies are denoted by a $C$ in the upper right portion of the spectrum.  See \citetalias{butcher2016} for codes used in the previous campaigns.}
\label{fig:DET1}
\end{figure}

Color SDSS images along with Arecibo \HI\ line spectra of all 151 galaxies from the final follow-up \HI\ observing campaign are shown in Figures \ref{fig:DET1} to \ref{fig:DET3} for Arecibo detections, in Figure \ref{fig:MAR1} for marginal detections, and in Figures \ref{fig:ND1} and \ref{fig:ND2} for nondetections. 
Selected properties of the three categories of galaxy detections are listed in Tables \ref{tab:A3060_DET}, \ref{tab:A3060_MAR} and \ref{tab:A3060_ND} respectively. 

Listed throughout the Tables are the following properties of the target galaxies:

\begin{itemize}
\item{source:} NIBLES sample source number (see \citetalias{vandriel2016}); 
\item{RA \& Dec:} Right Ascension and Declination in J2000.0 coordinates, as used for the observations;
\item{Name:} common catalog name, other than the SDSS;
\item{\Vopt:} heliocentric recession velocity from the SDSS redshift, determined in the optical convention (in \kms), from \citetalias{butcher2016};
\item{$g-z$:} $g-z$ extinction-corrected (following \cite{schlegel1998}) color of the galaxy using SDSS model magnitudes;
\item{\Mg:} extinction corrected absolute $g$-band magnitude;
\item{log(\Mstar/\Msun):} total median stellar mass estimates;
\item{log(\Lr/\Lsun):} SDSS $r$-band luminosity derived from Petrosian magnitudes as in \citetalias{butcher2018}; 
\item{log(sSFR/yr$^{-1}$):} specific Star Formation Rate, or SFR/\Mstar;
\item{$rms$:} rms noise level values of the \HI\ spectra (in mJy);
\item{\VHI:} heliocentric recession velocity of the center of the \HI\ line profile (in \kms);
\item{\Wfifty, \Wtwenty:} velocity widths measured at 50\% and 20\% of the \HI\ profile peak level, 
respectively, uncorrected for galaxy inclination (in \kms);
\item{\FHI:} integrated \HI\ line flux (in \Jykms);
\item{$SNR$:} peak signal-to-noise ratio, which we define as the peak flux density divided by the rms.
For nondetections, the $SNR$ listed is the maximum found in the expected velocity range of the \HI\ profile; 
\item{$S/N$:} signal-to-noise, determined taking into account the line width, following the ALFALFA \HI\ survey formulation from \citet{saintonge07}: 
$S/N$ = 1000(\FHI/\Wfifty)$\cdot$(\Wfifty/2$\cdot$R)$^{0.5}$)/rms, where $R$ is the velocity resolution, 18.7 \kms;
\item{log(\MHI/\Msun):} Total \HI\ mass, where \MHI\ = 2.36$\times$$10^5$$\cdot$$D^2$$\cdot$\FHI, where $D$ = $V$/70 is the galaxy's distance (in Mpc). In the cases of nondetections, 3$\sigma$ upper limits are listed for a flat-topped profile with a width depending on the target’s $r$-band luminosity, \Lr, according to the upper envelope in the \Wtwenty\ - log(\Lr/\Lsun) relationship of our \nan\ clear, nonconfused detections (see \citetalias{vandriel2016}) - these are quite conservative upper limits;
\item{log(\MHI/\Mstar):} ratio of the total \HI\ and stellar masses (in \Msun).
\end{itemize}

Estimated uncertainties are given after the values in the tables. Uncertainties in the central \HI\ line velocity, \VHI, and in the integrated \HI\ line flux, \FHI, were determined following \citet{schneider86, schneider90} as, respectively 
\begin{equation} 
\sigma_{v_{HI}} = 1.5(W_{20}-W_{50})S\!NR^{-1}\, (\kms)
\end{equation} 
and
\begin{equation} 
\sigma_{F_{HI}} = 2(1.2W_{20}R)^{0.5}rms\, (\textrm{Jy}\ \kms)
\end{equation} 
where $R$ is the instrumental resolution, 18.7 \kms, $SNR$ is the peak signal-to-noise ratio of a spectrum and $rms$ is the rms noise level (in Jy). Following Schneider et al., the uncertainty in the \Wfifty\ and \Wtwenty\ line widths is expected to be 2 and 3.1 times the uncertainty in \VHI, respectively.  


\onecolumn

\begin{figure} 
\centering
\includegraphics[width=15cm]{DET2.pdf}
\caption{\label{DET2} Color images from the SDSS alongside the 21-cm \HI\ line spectra of galaxies clearly detected at Arecibo  -- {\it continued}. See Figure \ref{fig:DET1} for further details. }
\label{fig:DET2}
\end{figure}

\begin{figure} 
\centering
\includegraphics[width=15cm]{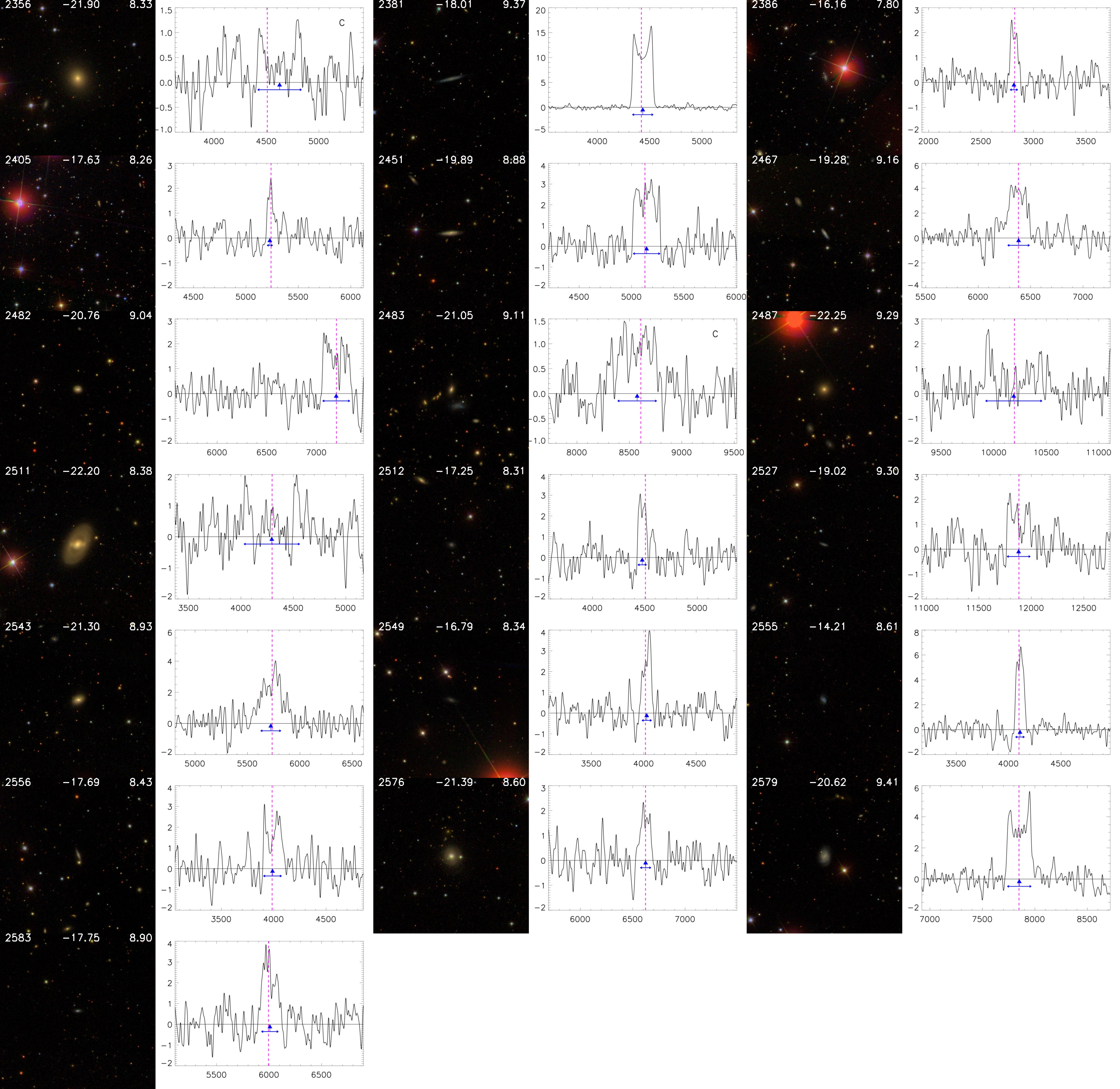}
\caption{\label{DET3} Color images from the SDSS alongside the 21-cm \HI\ line spectra of galaxies clearly detected at Arecibo  -- {\it continued}. See Figure \ref{fig:DET1} for further details. }
\label{fig:DET3}
\end{figure}

\onecolumn
\begin{figure} 
\centering
\includegraphics[width=15cm]{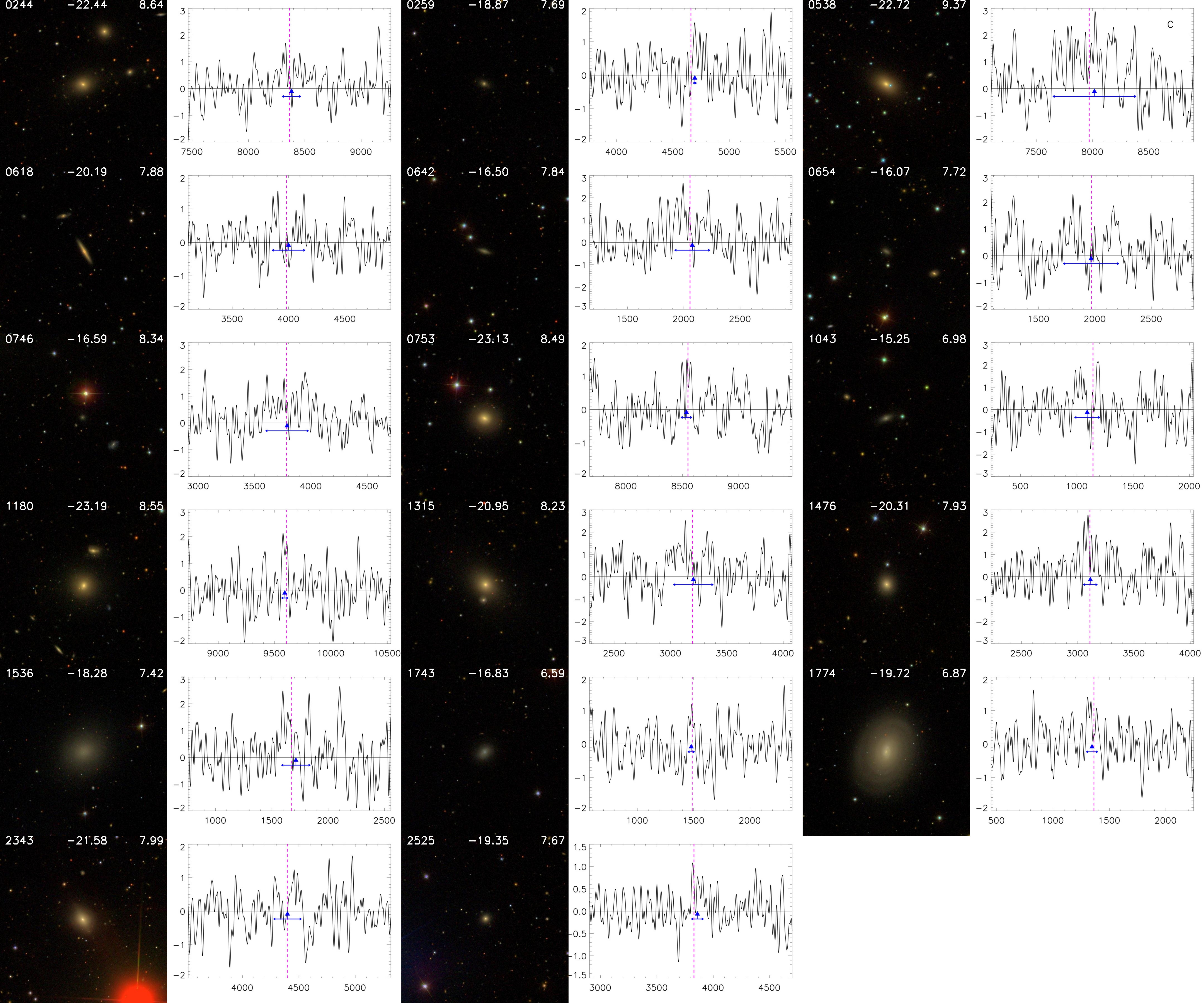}
\caption{\label{MAR1} Color images from the SDSS alongside the 21-cm \HI\ line spectra of galaxies marginally detected at Arecibo. See Figure \ref{fig:DET1} for further details. }
\label{fig:MAR1}
\end{figure}

\onecolumn
\begin{figure} 
\centering
\includegraphics[width=15cm]{ND1.pdf}
\caption{\label{ND1} Color images from the SDSS alongside the 21-cm \HI\ line spectra of galaxies undetected at Arecibo. See Figure \ref{fig:DET1} for further details. }
\label{fig:ND1}
\end{figure}

\onecolumn
\begin{figure} 
\centering
\includegraphics[width=15cm]{ND2.pdf}
\caption{\label{ND2} Color images from the SDSS alongside the 21-cm \HI\ line spectra of galaxies undetected at Arecibo   -- {\it continued}. See Figure \ref{fig:DET1} for further details. }
\label{fig:ND2}
\end{figure}


\onecolumn
{\fontsize{7}{7}\selectfont
\begin{landscape}
\begin{longtable}{lrrrrrrrrrrrrrrrrrr}  
\caption{\label{A3060_DET} Basic optical and \HI\ data -- Arecibo detections of \nan\ nondetections and marginals} \\
\hline \hline
Source & RA & DEC                      & Name       & \Vopt\ & $g-z$ & \Mg\  & log(\Mstar) & log(\Lr)   & log(\fsSFR)   & $rms$ & \VHI\ & \Wfifty\ & \Wtwenty\ & \FHI\   & $SNR$ & $S/N$ & log(\fMHIMsun)  & log(\fMHIMstar) \\ 
       & \multicolumn{2}{c}{(J2000.0)} &            &        &                 &    &   &             &       &          &           &         &       &       &   &   &  &   \\
       &    &                          &            & km/s  & mag     & mag   &   &  &  & mJy   & km/s  & km/s    & km/s       & Jy km/s &       &       &  &              \\
\hline
\endfirsthead
\caption{\it continued.} \\
\hline \hline
Source & RA & DEC                      & Name       & \Vopt\ & $g-z$ & \Mg\  & log(\Mstar) & log(\Lr)   & log(\fsSFR)   & $rms$ & \VHI\ & \Wfifty\ & \Wtwenty\ & \FHI\   & $SNR$ & $S/N$ & log(\fMHIMsun)  & log(\fMHIMstar) \\ 
       & \multicolumn{2}{c}{(J2000.0)} &            &        &                 &    &   &             &       &          &           &         &       &       &   &   &  &   \\
       &    &                          &            & km/s  & mag     & mag   &   &  &  & mJy   & km/s  & km/s    & km/s       & Jy km/s &       &       &  &              \\
\hline 
\endhead
\hline
\endfoot
0003        & 00 00 47.90 &  14 16 39.10  & \object{2MASX J00004789+1416390} & 10847$\pm$3  &  1.66 & -20.01 & 10.71 & 10.06 &   -10.49 & 0.43 & 10843$\pm$12 & 369 &  413 &  0.38$\pm$0.08 &   5.1 &    8.1 &  9.34 & -1.37 \\ 
0015        & 00 06 29.29 &  14 10 56.40  & \object{2MASX J00062933+1410559} &  5324$\pm$6  &  1.59 & -17.84 &  9.74 &  9.19 &   -11.30 & 0.51 &  5311$\pm$3  & 262 &  279 &  0.67$\pm$0.08 &   8.0 &   13.2 &  8.96 & -0.78 \\ 
0018        & 00 08 00.98 &  14 01 18.60  & \object{             ASK 146993} &  5429$\pm$4  &  0.76 & -17.38 &  8.74 &  8.72 &    -9.73 & 0.55 &  5435$\pm$4  & 142 &  166 &  0.59$\pm$0.07 &   8.3 &   14.5 &  8.92 &  0.18 \\ 
0024$^M$    & 00 10 52.65 &  15 35 26.60  & \object{2MASX J00105266+1535261} & 11071$\pm$6  &  2.19 & -20.06 & 11.29 & 10.17 &   -11.27 & 0.47 & 11261$\pm$44 &  98 &  458 &  1.44$\pm$0.10 &  12.1 &   49.4 &  9.94 & -1.35 \\ 
0025        & 00 11 43.20 &  14 28 01.10  & \object{             ASK 146943} &  5213$\pm$2  &  0.66 & -17.47 &  8.67 &  8.91 &    -9.48 & 0.39 &  5203$\pm$8  &  85 &  123 &  0.19$\pm$0.04 &   6.6 &    8.7 &  8.40 & -0.27 \\ 
0026        & 00 11 44.80 &  14 32 13.60  & \object{             ASK 146940} &  5121$\pm$2  &  0.64 & -16.57 &  8.23 &  8.54 &    -9.50 & 0.30 &  5122$\pm$3  &  52 &   60 &  0.04$\pm$0.02 &   3.8 &    3.1 &  7.72 & -0.51 \\ 
0034        & 00 14 17.63 &  15 51 06.30  & \object{             ASK 147688} &  1943$\pm$5  &  0.44 & -14.52 &   --- &  7.70 &      --- & 0.60 &  1946$\pm$4  &  49 &   73 &  0.24$\pm$0.05 &   7.5 &    9.2 &  7.64 &   --- \\ 
0059        & 00 25 17.30 &  14 34 40.80  & \object{             ASK 147884} &  5224$\pm$1  &  0.44 & -16.59 &  8.11 &  8.59 &    -9.47 & 0.69 &  5240$\pm$6  & 126 &  143 &  0.23$\pm$0.08 &   3.6 &    4.7 &  8.48 &  0.37 \\ 
0108$^M$    & 00 41 42.50 &  14 17 56.20  & \object{              ASK 40372} &  8810$\pm$1  &  0.61 & -18.16 &  6.66 &  9.14 &    -8.59 & 0.81 &  8828$\pm$5  & 100 &  114 &  0.18$\pm$0.08 &   3.8 &    3.4 &  8.82 &  2.16 \\ 
0189        & 01 20 12.10 &  14 33 40.60  & \object{              UGC 00865} & 11332$\pm$4  &  1.68 & -21.44 & 11.36 & 10.76 &   -12.72 & 0.51 & 11384$\pm$12 & 286 &  315 &  0.30$\pm$0.09 &   3.6 &    5.5 &  9.27 & -2.09 \\ 
0221$^M$    & 01 31 01.08 &  13 03 15.60  & \object{2MASX J01310103+1303155} &  6206$\pm$2  &  1.45 & -19.38 & 10.28 &  9.87 &   -12.20 & 0.87 &  6213$\pm$38 & 200 &  311 &  0.51$\pm$0.15 &   4.3 &    6.7 &  8.98 & -1.30 \\ 
0223        & 01 31 26.60 &  14 06 15.10  & \object{               SHOC 071} &  4155$\pm$0  &  0.56 & -17.11 &  8.45 &  8.72 &    -9.52 & 1.04 &  4156$\pm$4  &  42 &   51 &  0.12$\pm$0.07 &   3.4 &    2.8 &  7.98 & -0.47 \\ 
0231$^M$    & 01 33 52.56 &  13 42 09.40  & \object{              ASK 43205} &  2599$\pm$1  & -0.57 & -15.14 &  6.60 &  7.81 &    -8.04 & 0.64 &  2592$\pm$6  &  33 &   58 &  0.15$\pm$0.05 &   6.2 &    6.6 &  7.69 &  1.09 \\ 
0253        & 01 46 27.70 &  12 53 34.20  & \object{            PGC 4124972} &  5166$\pm$13 &  0.70 & -17.35 &  6.47 &  8.92 &   -10.23 & 0.57 &  5158$\pm$2  &  35 &   50 &  0.27$\pm$0.04 &  11.0 &   13.0 &  8.54 &  2.07 \\ 
0478$^M$    & 07 36 07.60 &  30 12 54.20  & \object{2MASX J07360760+3012543} &  6979$\pm$2  &  0.88 & -18.97 &  9.56 &  9.58 &    -9.98 & 1.13 &  6958$\pm$54 & 229 &  339 &  0.53$\pm$0.20 &   3.1 &    5.0 &  9.09 & -0.46 \\ 
0484        & 07 36 59.50 &  30 29 53.30  & \object{            PGC 1906163} &  4546$\pm$2  &  0.72 & -16.86 &  8.54 &  8.70 &    -9.76 & 0.60 &  4559$\pm$15 &  61 &  102 &  0.14$\pm$0.06 &   3.8 &    4.8 &  8.15 & -0.39 \\ 
0492$^U$    & 07 38 26.50 &  28 57 47.00  & \object{           CGCG 147-047} &  4732$\pm$3  &  1.41 & -18.47 &  9.83 &  9.54 &   -11.06 & 0.74 &  4733$\pm$3  & 282 &  294 &  0.71$\pm$0.12 &   5.9 &    9.2 &  8.88 & -0.94 \\ 
0506        & 07 40 22.70 &  23 16 29.90  & \object{              UGC 03960} &  2255$\pm$2  &  1.23 & -18.66 &  9.77 &  9.54 &   -11.59 & 0.61 &  2288$\pm$14 & 137 &  193 &  0.41$\pm$0.08 &   5.8 &    9.3 &  8.01 & -1.75 \\ 
0548        & 07 54 10.20 &  26 11 00.70  & \object{             ASK 203882} &  4661$\pm$1  &  0.31 & -17.02 &  6.57 &  8.70 &    -9.49 & 0.51 &  4675$\pm$8  &  52 &   94 &  0.22$\pm$0.05 &   7.4 &    9.9 &  8.37 &  1.81 \\ 
0584        & 08 05 47.80 &  30 13 52.30  & \object{           CGCG 148-111} &  2306$\pm$3  &  0.98 & -17.11 &  9.17 &  8.90 &      --- & 0.56 &  2306$\pm$7  & 116 &  156 &  0.41$\pm$0.07 &   7.5 &   11.0 &  8.02 & -1.15 \\ 
0606$^M$    & 08 16 57.60 &  20 30 44.10  & \object{           CGCG 119-027} &  4334$\pm$4  &  1.04 & -18.79 &  9.62 &  9.57 &   -10.63 & 0.51 &  4328$\pm$3  & 166 &  191 &  0.57$\pm$0.07 &   9.5 &   14.1 &  8.71 & -0.91 \\ 
0608        & 08 17 15.90 &  24 53 56.80  & \object{ 2MASXJ08171594+2453569} &  1898$\pm$7  &  1.15 & -15.42 &  8.42 &  8.30 &   -11.01 & 0.77 &  1900$\pm$3  & 342 &  350 &  0.32$\pm$0.14 &   3.4 &    3.7 &  7.75 & -0.67 \\ 
0623        & 08 19 32.10 &  21 23 39.40  & \object{                IC 2293} &  4090$\pm$3  &  1.23 & -19.35 &  9.98 &  9.85 &   -10.82 & 0.58 &  4094$\pm$5  & 220 &  238 &  0.48$\pm$0.09 &   5.3 &    9.1 &  8.59 & -1.39 \\ 
0628        & 08 20 23.70 &  21 07 53.20  & \object{               NGC 2562} &  5062$\pm$1  &  1.54 & -20.97 & 10.96 & 10.53 &   -12.11 & 1.02 &  5075$\pm$8  & 580 &  599 &  0.41$\pm$0.24 &   3.4 &    2.7 &  8.71 & -2.25 \\ 
0637        & 08 22 23.10 &  22 41 48.60  & \object{2MASX J08222312+2241484} &  2077$\pm$3  &  1.04 & -16.64 &  8.72 &  8.56 &   -10.42 & 0.73 &  2068$\pm$10 &  25 &   56 &  0.12$\pm$0.05 &   4.7 &    5.4 &  7.39 & -1.32 \\ 
0656        & 08 26 33.80 &  25 29 59.20  & \object{           KUG 0823+256} &  2148$\pm$2  &  0.90 & -15.47 &  8.16 &  8.17 &    -9.91 & 0.83 &  2151$\pm$3  & 105 &  114 &  0.22$\pm$0.08 &   3.3 &    4.2 &  7.69 & -0.47 \\ 
0657        & 08 26 39.20 &  25 35 53.50  & \object{2MASX J08263919+2535534} &  2332$\pm$2  &  1.17 & -16.48 &  8.81 &  8.50 &   -10.21 & 0.47 &  2337$\pm$15 & 104 &  135 &  0.13$\pm$0.05 &   3.1 &    4.4 &  7.53 & -1.28 \\ 
0688        & 08 36 28.40 &  18 21 38.90  & \object{           CGCG 089-052} &  4006$\pm$2  &  0.93 & -18.24 &  9.28 &  9.35 &   -10.07 & 0.73 &  4048$\pm$10 & 116 &  153 &  0.29$\pm$0.09 &   5.2 &    6.1 &  8.37 & -0.91 \\ 
0695        & 08 38 10.55 &  18 41 30.30  & \object{           CGCG 089-054} &  5859$\pm$1  &  1.46 & -19.76 & 10.43 & 10.00 &   -11.76 & 0.94 &  5860$\pm$20 &  30 &   74 &  0.11$\pm$0.08 &   3.3 &    3.4 &  8.26 & -2.17 \\ 
0730        & 08 46 36.00 &  19 00 50.00  & \object{              UGC 04588} &  4377$\pm$2  &  1.12 & -18.54 &   --- &  9.50 &      --- & 0.52 &  4262$\pm$1  & 272 &  287 &  4.06$\pm$0.08 &  17.6 &   77.2 &  9.55 &   --- \\ 
0737        & 08 47 49.72 &  18 54 42.80  & \object{                IC 2399} &  4225$\pm$7  &  0.97 & -18.29 &  9.44 &  9.38 &   -10.57 & 0.82 &  4222$\pm$10 &  71 &   96 &  0.19$\pm$0.08 &   3.5 &    4.3 &  8.21 & -1.23 \\ 
0738        & 08 47 59.49 &  19 02 56.40  & \object{            PGC 1578852} &  3844$\pm$2  &  0.87 & -16.66 &  8.50 &  8.71 &   -10.05 & 0.67 &  3843$\pm$6  & 137 &  159 &  0.36$\pm$0.08 &   5.0 &    7.5 &  8.41 & -0.09 \\ 
0744        & 08 49 01.40 &  29 29 18.20  & \object{           KUG 0845+296} &  8239$\pm$1  & -0.03 & -17.60 &  6.80 &  8.92 &    -8.72 & 0.68 &  8234$\pm$4  &  59 &   79 &  0.30$\pm$0.06 &   7.3 &    9.2 &  8.99 &  2.19 \\ 
0748$^U$    & 08 49 24.19 &  19 04 27.50  & \object{               NGC 2673} &  3749$\pm$1  &  1.48 & -19.64 & 10.35 & 10.26 &   -12.86 & 0.93 &  3745$\pm$4  & 404 &  414 &  0.30$\pm$0.18 &   3.3 &    2.6 &  8.31 & -2.05 \\ 
0917$^M$    & 09 43 31.10 &  31 58 37.10  & \object{               NGC 2970} &  1630$\pm$2  &  1.06 & -17.93 &  9.17 &  9.17 &   -10.56 & 0.54 &  1650$\pm$9  &  50 &  135 &  0.61$\pm$0.06 &  13.2 &   25.9 &  7.90 & -1.27 \\ 
1021$^M$    & 10 17 39.70 &  22 48 24.30  & \object{           CGCG 123-035} &  1260$\pm$3  &  1.23 & -16.05 &   --- &  8.54 &      --- & 0.53 &  1180$\pm$2  & 187 &  205 &  1.09$\pm$0.07 &  11.1 &   24.5 &  7.86 &   --- \\ 
1132        & 10 56 19.90 &  17 05 05.90  & \object{            PGC 4257755} &   960$\pm$1  &  0.44 & -12.53 &  6.57 &  7.03 &    -9.47 & 0.53 &   967$\pm$4  &  23 &   33 &  0.03$\pm$0.03 &   3.1 &    2.0 &  6.16 & -0.41 \\ 
1134$^M$    & 10 56 38.70 &  17 23 01.20  & \object{            PGC 4573180} &   948$\pm$1  &  0.50 & -12.81 &  6.63 &  7.12 &    -9.35 & 0.59 &   944$\pm$2  &  23 &   34 &  0.09$\pm$0.03 &   6.0 &    5.2 &  6.59 & -0.04 \\ 
1155        & 11 03 26.30 &  16 00 58.20  & \object{           LSBC F640-02} &  1235$\pm$14 &  0.86 & -14.66 &  7.74 &  7.93 &   -10.19 & 0.62 &  1226$\pm$2  &  98 &  111 &  0.57$\pm$0.06 &   9.7 &   15.3 &  7.61 & -0.13 \\ 
1191        & 11 13 10.40 &  27 49 05.60  & \object{              UGC 06250} &  9367$\pm$2  &  1.59 & -21.66 & 11.30 & 10.82 &   -12.67 & 0.53 &  9395$\pm$4  & 563 &  583 &  1.49$\pm$0.13 &   7.0 &   18.7 &  9.80 & -1.50 \\ 
1534$^M$    & 12 09 09.80 &  31 34 10.20  & \object{              UGC 07132} &  6766$\pm$3  &  1.45 & -21.31 & 11.02 & 10.69 &   -12.62 & 0.75 &  6806$\pm$11 & 195 &  228 &  0.37$\pm$0.11 &   4.2 &    5.7 &  8.92 & -2.10 \\ 
1556        & 12 12 11.80 &  13 14 47.40  & \object{               NGC 4165} &  1868$\pm$4  &  1.16 & -18.22 &  9.49 &  9.41 &   -11.42 & 0.79 &  1873$\pm$17 & 206 &  285 &  1.01$\pm$0.13 &   6.7 &   14.6 &  8.23 & -1.25 \\ 
1591        & 12 17 08.20 &  27 47 42.90  & \object{            PGC 1818175} &   971$\pm$10 &  0.79 & -14.88 &   --- &  8.00 &      --- & 0.58 &   967$\pm$3  &  33 &   45 &  0.08$\pm$0.04 &   4.8 &    4.1 &  6.58 &   --- \\ 
1696        & 12 38 17.90 &  13 06 35.60  & \object{               NGC 4584} &  1715$\pm$3  &  1.04 & -18.35 &  9.46 &  9.44 &   -11.07 & 0.62 &  1731$\pm$9  &  47 &   83 &  0.18$\pm$0.05 &   5.6 &    7.0 &  7.42 & -2.03 \\ 
1703$^M$    & 12 39 48.00 &  12 58 26.20  & \object{                IC 3631} &  2828$\pm$3  &  0.73 & -19.31 &  9.50 &  9.70 &    -9.84 & 0.46 &  2805$\pm$5  & 302 &  325 &  0.42$\pm$0.08 &   6.5 &    8.5 &  8.20 & -1.30 \\ 
1726        & 12 47 16.10 &  11 45 36.70  & \object{                IC 3775} &  1093$\pm$13 &  0.86 & -14.76 &   --- &  7.93 &      --- & 0.54 &  1073$\pm$4  &  27 &   37 &  0.04$\pm$0.03 &   3.2 &    2.1 &  6.31 &   --- \\ 
1733        & 12 48 55.40 &  14 54 28.30  & \object{                IC 3806} &  1401$\pm$5  &  1.20 & -17.59 &  9.31 &  9.12 &   -11.95 & 0.51 &  1412$\pm$9  & 140 &  170 &  0.28$\pm$0.06 &   4.9 &    7.6 &  7.43 & -1.87 \\ 
1744        & 12 51 55.10 &  12 05 00.70  & \object{               NGC 4746} &  1744$\pm$4  &  1.46 & -18.51 &   --- &  9.62 &      --- & 0.83 &  1779$\pm$1  & 330 &  355 & 11.78$\pm$0.15 &  18.5 &  127.8 &  9.25 &   --- \\ 
1768$^M$    & 12 59 06.50 &  28 48 42.40  & \object{             ASK 511090} &   999$\pm$7  &  0.49 & -13.81 &  7.19 &  7.46 &    -9.70 & 0.46 &   983$\pm$2  &  28 &   41 &  0.12$\pm$0.03 &   8.2 &    8.0 &  6.74 & -0.45 \\ 
1893        & 13 38 43.10 &  31 16 13.90  & \object{           CGCG 161-101} &  4699$\pm$2  &  1.53 & -19.85 & 10.52 & 10.06 &   -12.39 & 0.66 &  4728$\pm$16 & 259 &  299 &  0.36$\pm$0.11 &   3.6 &    5.5 &  8.59 & -1.93 \\ 
1912        & 13 44 23.20 &  05 47 32.60  & \object{              UGC 08689} &  6840$\pm$2  &  1.49 & -21.36 & 11.00 & 10.63 &   -12.08 & 0.61 &  6859$\pm$20 & 379 &  432 &  0.40$\pm$0.12 &   3.8 &    5.4 &  8.96 & -2.04 \\ 
1946        & 13 52 05.60 &  05 45 54.10  & \object{            PGC 4389504} &   984$\pm$2  &  0.71 & -13.95 &  7.52 &  7.64 &   -10.54 & 0.46 &   987$\pm$2  &  30 &   41 &  0.08$\pm$0.03 &   6.1 &    5.4 &  6.59 & -0.92 \\ 
1949        & 13 52 10.80 &  05 30 13.80  & \object{             ASK 178807} &  1216$\pm$19 &  0.61 & -14.27 &  6.09 &  7.72 &   -10.45 & 0.40 &  1233$\pm$2  &  66 &   78 &  0.22$\pm$0.03 &   8.7 &   11.0 &  7.21 &  1.12 \\ 
1992        & 13 58 12.80 &  06 31 05.40  & \object{               NGC 5384} &  5094$\pm$1  &  1.56 & -20.74 & 10.88 & 10.43 &   -11.68 & 0.83 &  5046$\pm$14 & 217 &  251 &  0.32$\pm$0.13 &   3.5 &    4.3 &  8.59 & -2.29 \\ 
2152        & 14 30 07.20 &  08 42 16.00  & \object{             ASK 457251} &  1427$\pm$35 &  0.53 & -13.36 &   --- &  7.36 &      --- & 0.62 &  1423$\pm$6  &  23 &   37 &  0.05$\pm$0.04 &   3.2 &    2.8 &  6.70 &   --- \\ 
2230        & 14 47 00.10 &  11 35 31.20  & \object{              UGC 09521} &  9343$\pm$2  &  1.56 & -21.66 & 11.22 & 10.83 &   -12.32 & 0.75 &  9353$\pm$8  & 500 &  526 &  1.05$\pm$0.17 &   4.9 &    9.9 &  9.64 & -1.58 \\ 
2305        & 15 21 55.30 &  08 25 25.60  & \object{                IC 1116} & 11706$\pm$3  &  1.50 & -22.18 & 11.41 & 10.99 &   -12.28 & 0.75 & 11702$\pm$14 & 255 &  288 &  0.34$\pm$0.12 &   3.4 &    4.6 &  9.36 & -2.05 \\ 
2337        & 15 42 35.50 &  23 40 12.10  & \object{                IC 4576} &  6741$\pm$2  &  1.47 & -21.04 & 10.89 & 10.52 &   -12.47 & 0.47 &  6704$\pm$5  & 351 &  364 &  0.14$\pm$0.09 &   3.6 &    2.6 &  8.48 & -2.41 \\ 
2342        & 15 56 14.40 &  06 05 53.40  & \object{             ASK 461670} &  1797$\pm$3  &  1.41 & -11.64 &   --- &  6.56 &      --- & 0.36 &  1788$\pm$2  &  33 &   50 &  0.19$\pm$0.02 &  11.6 &   14.7 &  7.46 &   --- \\ 
2353        & 16 02 11.60 &  07 05 09.90  & \object{           CGCG 051-021} &  2654$\pm$2  &  1.45 & -18.69 &  9.98 &  9.55 &   -10.27 & 0.40 &  2647$\pm$6  &  98 &  121 &  0.18$\pm$0.04 &   4.9 &    7.4 &  7.79 & -2.19 \\ 
2356$^{CM}$ & 16 02 30.50 &  21 07 14.50  & \object{           CGCG 137-019} &  4507$\pm$2  &  1.48 & -20.42 & 10.65 & 10.20 &   -12.30 & 0.38 &  4625$\pm$4  & 417 &  427 &  0.21$\pm$0.08 &   3.5 &    4.3 &  8.33 & -2.33 \\ 
2381        & 16 46 54.60 &  31 53 07.80  & \object{               FGC 2068} &  4419$\pm$7  &  0.56 & -17.45 &  6.74 &  8.96 &   -10.73 & 0.34 &  4432$\pm$1  & 197 &  213 &  2.46$\pm$0.05 &  18.5 &   84.4 &  9.37 &  2.63 \\ 
2386        & 17 01 29.60 &  22 43 52.40  & \object{             ASK 406590} &  2820$\pm$1  &  0.75 & -15.41 &  8.11 &  8.13 &    -9.92 & 0.36 &  2813$\pm$6  &  73 &  102 &  0.17$\pm$0.03 &   6.6 &    8.8 &  7.80 & -0.31 \\ 
2405        & 21 15 16.20 &  09 53 47.10  & \object{             ASK 138334} &  5239$\pm$1  &  0.49 & -17.14 &  8.38 &  8.78 &    -9.40 & 0.45 &  5228$\pm$15 &  52 &  105 &  0.14$\pm$0.04 &   5.2 &    6.9 &  8.26 & -0.12 \\ 
2451        & 22 24 44.10 &  12 21 40.70  & \object{2MASX J22244417+1221410} &  5127$\pm$3  &  1.35 & -18.54 &  9.80 &  9.49 &   -10.43 & 0.57 &  5143$\pm$2  & 255 &  265 &  0.60$\pm$0.09 &   5.5 &   10.6 &  8.88 & -0.92 \\ 
2467        & 22 40 17.40 &  14 29 56.80  & \object{2MASX J22401742+1429556} &  6384$\pm$1  &  0.70 & -18.57 &  9.20 &  9.36 &    -9.64 & 0.67 &  6385$\pm$6  & 207 &  231 &  0.73$\pm$0.10 &   5.8 &   12.2 &  9.16 & -0.04 \\ 
2482        & 22 58 53.20 &  13 21 31.40  & \object{           KUG 2256+130} &  7200$\pm$3  &  1.22 & -19.54 & 10.08 &  9.86 &   -10.10 & 0.47 &  7198$\pm$4  & 273 &  288 &  0.44$\pm$0.08 &   4.8 &    9.1 &  9.04 & -1.05 \\ 
2483$^{CM}$ & 22 59 00.20 &  13 29 51.60  & \object{2MASX J22590020+1329513} &  8609$\pm$1  &  1.24 & -19.81 & 10.21 &  9.90 &   -10.30 & 0.37 &  8574$\pm$8  & 371 &  392 &  0.36$\pm$0.07 &   3.7 &    8.0 &  9.11 & -1.10 \\ 
2487        & 22 59 36.30 &  13 22 11.40  & \object{2MASX J22593625+1322112} & 10195$\pm$3  &  1.57 & -20.69 & 10.86 & 10.35 &   -12.38 & 0.51 & 10189$\pm$3  & 537 &  551 &  0.39$\pm$0.12 &   5.4 &    5.2 &  9.29 & -1.57 \\ 
2511        & 23 15 55.90 &  13 11 46.00  & \object{               NGC 7563} &  4297$\pm$2  &  1.52 & -20.69 & 10.83 & 10.44 &   -12.51 & 0.63 &  4294$\pm$5  & 531 &  542 &  0.27$\pm$0.14 &   3.3 &    3.0 &  8.38 & -2.45 \\ 
2512        & 23 16 03.80 &  13 47 05.60  & \object{             ASK 143916} &  4504$\pm$6  &  0.94 & -16.32 &  8.38 &  8.51 &    -9.90 & 0.59 &  4474$\pm$4  &  82 &   97 &  0.21$\pm$0.06 &   5.0 &    6.3 &  8.31 & -0.08 \\ 
2527$^M$    & 23 21 24.70 &  14 11 52.90  & \object{            PGC 4125637} & 11881$\pm$2  &  0.72 & -18.30 &  7.24 &  9.27 &   -10.74 & 0.60 & 11879$\pm$3  & 221 &  230 &  0.29$\pm$0.09 &   3.7 &    5.2 &  9.30 &  2.06 \\ 
2543        & 23 33 27.00 &  14 20 06.80  & \object{           CGCG 432-002} &  5735$\pm$3  &  1.30 & -20.01 & 10.37 & 10.07 &   -10.54 & 0.59 &  5722$\pm$5  & 192 &  215 &  0.54$\pm$0.08 &   6.2 &   10.5 &  8.93 & -1.44 \\ 
2549$^M$    & 23 36 08.50 &  15 44 36.60  & \object{            PGC 4125646} &  4016$\pm$4  &  0.73 & -16.06 &   --- &  8.33 &      --- & 0.51 &  4028$\pm$3  &  91 &  110 &  0.28$\pm$0.05 &   7.3 &    9.3 &  8.34 &   --- \\ 
2555$^M$    & 23 37 37.10 &  15 08 26.90  & \object{           KUG 2335+148} &  4099$\pm$2  & -0.31 & -14.52 &  7.55 &  7.65 &    -9.87 & 0.46 &  4108$\pm$2  &  84 &  103 &  0.50$\pm$0.04 &  11.7 &   19.1 &  8.61 &  1.06 \\ 
2556$^M$    & 23 38 26.00 &  15 23 10.90  & \object{             ASK 145449} &  3984$\pm$3  &  1.02 & -16.67 &  8.78 &  8.72 &   -10.17 & 0.61 &  3988$\pm$9  & 172 &  203 &  0.35$\pm$0.08 &   5.0 &    7.1 &  8.43 & -0.36 \\ 
2576        & 23 47 03.78 &  14 50 30.30  & \object{           CGCG 432-030} &  6622$\pm$2  &  1.22 & -20.17 & 10.39 & 10.09 &   -10.75 & 0.55 &  6623$\pm$16 & 101 &  147 &  0.19$\pm$0.06 &   4.2 &    5.4 &  8.60 & -1.79 \\ 
2579$^M$    & 23 48 45.10 &  15 55 43.50  & \object{           KUG 2346+156} &  7848$\pm$10 &  0.96 & -19.66 &  9.93 &  9.89 &   -10.10 & 0.45 &  7851$\pm$4  & 227 &  257 &  0.86$\pm$0.07 &  10.4 &   20.2 &  9.41 & -0.52 \\ 
2583$^M$    & 23 51 14.64 &  15 22 23.80  & \object{             ASK 146293} &  5994$\pm$1  &  0.61 & -17.14 &  8.55 &  8.86 &    -9.64 & 0.71 &  6007$\pm$15 & 160 &  214 &  0.46$\pm$0.10 &   5.3 &    8.2 &  8.90 &  0.35 \\ 
\label{tab:A3060_DET}
\end{longtable}
\tablefoot{Flags in Col. 1: $C$ for targets clearly confused with another galaxy within the Arecibo telescope beam, $M$ for marginal \nan\ detections, and $U$ for unreliable photometry.}
\end{landscape}
}

\onecolumn
{\fontsize{7}{7}\selectfont
\begin{landscape}
\begin{longtable}{lrrrrrrrrrrrrrrrrrr}  
\caption{\label{A3060_MAR} Basic optical and \HI\ data -- Arecibo marginal detections of \nan\ nondetections and marginals} \\
\hline \hline
Source & RA & DEC                      & Name       & \Vopt\ & $g-z$ & \Mg\  & log(\Mstar) & log(\Lr)   & log(\fsSFR)   & $rms$ & \VHI\ & \Wfifty\ & \Wtwenty\ & \FHI\   & $SNR$ & $S/N$ & log(\fMHIMsun)  & log(\fMHIMstar) \\ 
& \multicolumn{2}{c}{(J2000.0)} &            &        &                 &    &   &             &       &          &           &         &       &       &   &   &  &   \\
&    &                          &            & km/s  & mag     & mag   &   &  &  & mJy   & km/s  & km/s    & km/s       & Jy km/s &       &       &  &              \\
\hline
\endfirsthead
\endhead
\hline
\endfoot
0244        & 01 38 54.76 &  15 01 17.70  & \object{           CGCG 437-008} &  8364$\pm$2  &  1.51 & -20.93 & 10.91 & 10.48 &   -12.20 & 0.49 &  8381$\pm$1  & 164 &  169 &  0.13$\pm$0.06 &   3.3 &    3.3 &  8.64 & -2.27 \\ 
0259        & 01 48 53.10 &  13 25 26.10  & \object{ 2MASXi J0148531+132526} &  4659$\pm$2  &  1.31 & -17.56 &  9.41 &  9.04 &   -10.39 & 0.62 &  4693$\pm$9  &  29 &   45 &  0.05$\pm$0.04 &   2.5 &    2.2 &  7.69 & -1.72 \\ 
0538$^{CM}$ & 07 50 25.19 &  30 13 32.40  & \object{           CGCG 148-046} &  7970$\pm$2  &  1.67 & -21.05 & 11.18 & 10.53 &   -12.61 & 0.79 &  8016$\pm$5  & 739 &  753 &  0.76$\pm$0.21 &   3.7 &    5.7 &  9.37 & -1.80 \\ 
0618        & 08 18 08.80 &  23 02 51.30  & \object{                IC 2269} &  3980$\pm$2  &  1.75 & -18.45 & 10.18 &  9.58 &   -11.95 & 0.51 &  3999$\pm$4  & 291 &  301 &  0.10$\pm$0.08 &   3.0 &    1.8 &  7.88 & -2.30 \\ 
0642        & 08 23 24.10 &  22 16 00.40  & \object{             ASK 484591} &  2055$\pm$13 &  1.07 & -15.43 &  8.31 &  8.23 &   -11.79 & 0.79 &  2075$\pm$8  & 312 &  331 &  0.34$\pm$0.14 &   3.2 &    3.9 &  7.84 & -0.47 \\ 
0654        & 08 26 28.80 &  26 04 29.10  & \object{             ASK 364075} &  1968$\pm$17 &  0.91 & -15.16 &  8.16 &  8.11 &   -11.04 & 0.85 &  1965$\pm$10 & 491 &  513 &  0.28$\pm$0.18 &   3.0 &    2.4 &  7.72 & -0.45 \\ 
0746        & 08 49 06.59 &  19 00 26.30  & \object{            PGC 4178363} &  3783$\pm$18 &  1.08 & -15.52 &  8.35 &  8.34 &   -11.82 & 0.52 &  3787$\pm$15 & 380 &  415 &  0.31$\pm$0.10 &   3.4 &    5.0 &  8.34 & -0.02 \\ 
0753        & 08 51 56.67 &  16 56 41.30  & \object{              UGC 04639} &  8549$\pm$2  &  1.54 & -21.59 & 11.18 & 10.75 &   -12.34 & 0.61 &  8535$\pm$6  &  98 &  108 &  0.09$\pm$0.06 &   2.5 &    2.4 &  8.49 & -2.68 \\ 
1043        & 10 23 15.40 &  20 10 40.50  & \object{             ASK 606279} &  1143$\pm$4  &  1.14 & -14.11 &   --- &  7.78 &      --- & 0.81 &  1091$\pm$12 & 224 &  245 &  0.17$\pm$0.12 &   2.6 &    2.2 &  6.98 &   --- \\ 
1180        & 11 10 55.90 &  28 32 37.60  & \object{               NGC 3558} &  9606$\pm$3  &  1.55 & -21.65 & 11.25 & 10.81 &   -12.52 & 0.63 &  9590$\pm$3  &  57 &   66 &  0.08$\pm$0.05 &   3.4 &    2.7 &  8.55 & -2.69 \\ 
1315        & 11 34 11.70 &  12 30 44.30  & \object{               NGC 3731} &  3195$\pm$2  &  1.38 & -19.57 & 10.31 &  9.92 &   -12.59 & 0.85 &  3202$\pm$44 & 353 &  435 &  0.34$\pm$0.17 &   2.8 &    3.5 &  8.23 & -2.08 \\ 
1476        & 11 59 49.60 &  30 50 39.90  & \object{                IC 2986} &  3108$\pm$2  &  1.40 & -18.91 & 10.00 &  9.65 &   -11.92 & 0.91 &  3112$\pm$5  & 124 &  135 &  0.18$\pm$0.10 &   3.0 &    3.0 &  7.93 & -2.07 \\ 
1536        & 12 09 22.20 &  13 59 32.70  & \object{                IC 3019} &  1675$\pm$14 &  0.77 & -17.51 &  9.07 &  9.04 &   -11.89 & 0.81 &  1713$\pm$5  & 255 &  266 &  0.19$\pm$0.13 &   3.2 &    2.4 &  7.42 & -1.65 \\ 
1743        & 12 51 41.50 &  13 05 41.80  & \object{           CGCG 071-059} &  1487$\pm$5  &  0.80 & -16.02 &  8.30 &  8.42 &   -10.23 & 0.58 &  1478$\pm$6  &  59 &   69 &  0.04$\pm$0.05 &   2.0 &    1.3 &  6.59 & -1.70 \\ 
1774        & 13 00 10.60 &  12 28 59.90  & \object{               NGC 4880} &  1362$\pm$3  &  1.05 & -18.67 &  9.51 &  9.59 &   -12.24 & 0.56 &  1345$\pm$11 & 101 &  119 &  0.09$\pm$0.06 &   2.5 &    2.5 &  6.87 & -2.64 \\ 
2343        & 15 56 33.70 &  21 17 20.70  & \object{           CGCG 137-004} &  4397$\pm$1  &  1.23 & -20.35 & 10.40 & 10.19 &   -11.70 & 0.54 &  4399$\pm$7  & 245 &  257 &  0.11$\pm$0.08 &   2.3 &    2.0 &  7.99 & -2.41 \\ 
2525        & 23 20 28.20 &  15 04 20.90  & \object{2MASX J23202822+1504211} &  3828$\pm$2  &  1.28 & -18.08 &  9.57 &  9.29 &   -11.23 & 0.33 &  3859$\pm$5  & 106 &  117 &  0.07$\pm$0.03 &   3.2 &    3.1 &  7.67 & -1.90 \\ 

\label{tab:A3060_MAR}
\end{longtable}
\tablefoot{Flags in Col. 1: $C$ for targets clearly confused with another galaxy within the Arecibo telescope beam, $M$ for marginal \nan\ detections, and $U$ for unreliable photometry.}
\end{landscape}
}

\onecolumn
{\fontsize{8}{8}\selectfont
\begin{landscape}
\begin{longtable}{lrrrrrrrrrrrr}  
\caption{\label{A3060_ND} Basic optical and \HI\ data -- Arecibo nondetections of \nan\ nondetections and marginals} \\
\hline \hline
Source & RA & DEC                      & Name       & \Vopt\ & $g-z$ & \Mg\ & log(\Mstar) & log(\fLrLsun)   & log(\fsSFR)    & $rms$ & log(\fMHIMsun)  & log(\fMHIMstar) \\ 
       & \multicolumn{2}{c}{(J2000.0)} &            &        &         &      &    &   &             &   &   &     \\
       &    &                          &            & km/s   & mag     & mag  &   &   &  & mJy   &  &              \\
\hline 
\endfirsthead
\caption{\it continued.} \\
\hline \hline
Source & RA & DEC                      & Name       & \Vopt\ & $g-z$ & \Mg\ & log(\Mstar) & log(\fLrLsun)   & log(\fsSFR)    & $rms$ & log(\fMHIMsun)  & log(\fMHIMstar) \\ 
       & \multicolumn{2}{c}{(J2000.0)} &            &        &         &      &    &   &             &   &   &     \\
       &    &                          &            & km/s   & mag     & mag  &  &   &  & mJy   &  &              \\
\hline 
\hline 
\endhead
\hline
\endfoot
0014        & 00 06 19.61 &  14 19 38.70 & \object{2MASX J00061957+1419389} &  5448$\pm$3  &  1.31 & -18.22 &  9.66 &  9.23 & -11.16 &  0.43 &  $<$7.54 & $<$-2.11 \\ 
0178        & 01 15 54.30 &  13 21 12.00 & \object{           CGCG 436-016} &  4184$\pm$2  &  1.46 & -19.35 & 10.24 &  9.87 & -12.12 &  0.81 &  $<$7.58 & $<$-2.66 \\ 
0222        & 01 31 04.80 &  13 38 58.80 & \object{           CGCG 436-067} &  5889$\pm$2  &  1.61 & -19.53 & 10.49 &  9.96 & -12.29 &  0.87 &  $<$7.92 & $<$-2.57 \\ 
0489        & 07 37 49.90 &  28 39 10.50 & \object{2MASX J07374992+2839102} &  4731$\pm$7  &  1.01 & -16.96 &  8.86 &  8.84 & -11.18 &  0.60 &  $<$7.55 & $<$-1.31 \\ 
0493        & 07 38 32.50 &  29 11 07.90 & \object{           CGCG 147-048} & 11784$\pm$4  &  1.45 & -21.09 & 10.91 & 10.50 & -12.24 &  0.74 &  $<$8.46 & $<$-2.45 \\ 
0507        & 07 40 27.50 &  30 48 20.00 & \object{2MASX J07402743+3048201} & 10503$\pm$2  &  1.56 & -20.69 & 10.87 & 10.38 & -12.26 &  0.61 &  $<$8.26 & $<$-2.61 \\ 
0564        & 07 59 29.70 &  27 01 35.10 & \object{               NGC 2492} &  6649$\pm$2  &  1.50 & -21.67 & 11.23 & 10.77 & -12.48 &  0.51 &  $<$7.79 & $<$-3.44 \\ 
0579        & 08 02 57.50 &  16 17 57.50 & \object{              UGC 04190} &  4853$\pm$2  &  1.53 & -20.44 & 10.77 & 10.30 & -12.62 &  0.51 &  $<$7.52 & $<$-3.25 \\ 
0586        & 08 06 13.40 &  17 42 23.60 & \object{               NGC 2522} &  4706$\pm$2  &  1.72 & -20.78 & 11.05 & 10.47 & -12.37 &  0.56 &  $<$7.54 & $<$-3.50 \\ 
0611        & 08 17 35.00 &  20 54 11.00 & \object{               NGC 2553} &  4670$\pm$1  &  1.48 & -20.38 & 10.64 & 10.27 & -12.35 &  0.77 &  $<$7.66 & $<$-2.98 \\ 
0622        & 08 19 24.28 &  21 00 12.80 & \object{2MASX J08192430+2100125} &  3906$\pm$4  &  1.28 & -18.05 &  9.67 &  9.25 & -12.38 &  0.51 &  $<$7.32 & $<$-2.35 \\ 
0624        & 08 19 36.04 &  21 14 28.90 & \object{2MASX J08193606+2114291} &  4566$\pm$8  &  1.27 & -17.13 &  9.17 &  9.00 & -11.62 &  0.58 &  $<$7.50 & $<$-1.66 \\ 
0626        & 08 19 51.90 &  20 59 05.90 & \object{               NGC 2560} &  4883$\pm$2  &  1.54 & -20.57 & 10.78 & 10.38 & -12.04 &  0.58 &  $<$7.56 & $<$-3.22 \\ 
0627        & 08 20 09.90 &  27 05 36.50 & \object{              UGC 04341} &  5869$\pm$2  &  1.62 & -20.72 & 10.92 & 10.45 & -11.46 &  0.58 &  $<$7.73 & $<$-3.19 \\ 
0630        & 08 20 35.70 &  21 04 04.00 & \object{               NGC 2563} &  4509$\pm$2  &  1.54 & -21.55 & 11.10 & 10.81 & -12.52 &  1.02 &  $<$7.75 & $<$-3.35 \\ 
0694        & 08 38 09.63 &  19 43 32.40 & \object{               NGC 2624} &  4134$\pm$1  &  1.38 & -19.49 & 10.21 &  9.90 & -10.62 &  0.73 &  $<$7.54 & $<$-2.67 \\ 
0733        & 08 46 49.20 &  28 10 16.70 & \object{                IC 2393} &  6297$\pm$2  &  1.51 & -21.41 & 11.09 & 10.66 & -12.35 &  0.52 &  $<$7.68 & $<$-3.42 \\ 
0747        & 08 49 15.01 &  19 11 27.30 & \object{            PGC 4572078} &  3827$\pm$11 &  1.22 & -16.06 &  8.71 &  8.38 & -11.30 &  0.52 &  $<$7.32 & $<$-1.39 \\ 
0756        & 08 54 40.70 &  20 35 00.90 & \object{           CGCG 120-050} &  3785$\pm$1  &  1.35 & -19.39 & 10.15 &  9.91 & -11.81 &  0.61 &  $<$7.38 & $<$-2.78 \\ 
0983        & 10 03 58.80 &  22 16 33.80 & \object{              UGC 05420} &  6120$\pm$2  &  1.49 & -20.94 & 10.93 & 10.53 & -12.64 &  0.54 &  $<$7.70 & $<$-3.23 \\ 
1038        & 10 21 21.10 &  24 20 29.00 & \object{              UGC 05591} & 11118$\pm$2  &  1.53 & -21.52 & 11.21 & 10.72 & -11.99 &  0.53 &  $<$8.25 & $<$-2.96 \\ 
1048        & 10 25 47.70 &  26 34 14.60 & \object{           CGCG 154-013} &  5023$\pm$2  &  1.44 & -20.35 & 10.62 & 10.25 & -12.04 &  0.81 &  $<$7.69 & $<$-2.92 \\ 
1076        & 10 36 38.40 &  14 10 15.90 & \object{               NGC 3300} &  3017$\pm$1  &  1.42 & -20.24 & 10.53 & 10.23 & -12.29 &  0.81 &  $<$7.31 & $<$-3.23 \\ 
1118        & 10 50 45.50 &  28 28 08.70 & \object{               NGC 3400} &  1412$\pm$1  &  1.36 & -18.09 &   --- &  9.35 &    --- &  0.81 &  $<$6.65 &      --- \\ 
1138        & 10 58 37.60 &  09 03 01.60 & \object{              UGC 06062} &  2623$\pm$2  &  1.46 & -19.20 & 10.18 &  9.80 & -12.22 &  0.59 &  $<$7.05 & $<$-3.14 \\ 
1160        & 11 04 56.80 &  17 38 30.50 & \object{            PGC 4263693} &   907$\pm$6  &  0.90 & -13.48 &  7.39 &  7.44 & -10.61 &  0.62 &  $<$6.08 & $<$-1.31 \\ 
1187        & 11 12 31.70 &  16 17 22.60 & \object{          LSBC F640-V04} &  1198$\pm$15 &  0.89 & -14.63 &   --- &  7.92 &    --- &  0.63 &  $<$6.40 &      --- \\ 
1295        & 11 31 42.00 &  28 09 12.90 & \object{               NGC 3713} &  6898$\pm$2  &  1.46 & -21.49 & 11.07 & 10.70 & -12.06 &  0.53 &  $<$7.79 & $<$-3.29 \\ 
1549        & 12 11 02.70 &  12 06 14.40 & \object{                IC 0767} &  1877$\pm$2  &  1.12 & -18.21 &  9.41 &  9.30 & -10.70 &  0.81 &  $<$6.87 & $<$-2.55 \\ 
1551        & 12 11 07.80 &  14 16 29.30 & \object{                IC 3032} &  1184$\pm$5  &  1.05 & -16.24 &   --- &  8.56 &    --- &  0.81 &  $<$6.49 &      --- \\ 
1561        & 12 12 18.90 &  15 28 59.10 & \object{               VCC 0050} &  1209$\pm$12 &  0.91 & -15.31 &   --- &  8.15 &    --- &  0.79 &  $<$6.51 &      --- \\ 
1575        & 12 15 08.50 &  14 58 18.70 & \object{               VCC 0137} &  1152$\pm$7  &  0.97 & -14.14 &   --- &  7.72 &    --- &  0.79 &  $<$6.40 &      --- \\ 
1612        & 12 19 47.60 &  30 20 20.70 & \object{               NGC 4272} &  8446$\pm$2  &  1.47 & -21.89 & 11.26 & 10.88 & -12.42 &  0.58 &  $<$8.06 & $<$-3.20 \\ 
1701        & 12 39 37.70 &  10 58 32.60 & \object{               VCC 1803} &  1355$\pm$21 &  1.01 & -15.27 &  8.24 &  8.18 & -11.23 &  0.62 &  $<$6.48 & $<$-1.75 \\ 
1702        & 12 39 42.80 &  13 36 26.80 & \object{               VCC 1806} &  1034$\pm$8  &  1.00 & -14.19 &   --- &  7.76 &    --- &  0.62 &  $<$6.24 &      --- \\ 
1708        & 12 40 13.40 &  12 52 29.10 & \object{                IC 3635} &  1560$\pm$8  &  1.07 & -16.20 &  8.69 &  8.58 & -11.76 &  0.46 &  $<$6.44 & $<$-2.25 \\ 
1713        & 12 41 39.40 &  12 14 50.60 & \object{                IC 3663} &   927$\pm$10 &  0.90 & -15.34 &   --- &  8.17 &    --- &  0.46 &  $<$6.02 &      --- \\ 
1715        & 12 41 46.10 &  11 29 19.10 & \object{                IC 3665} &  1227$\pm$29 &  0.81 & -16.16 &   --- &  8.50 &    --- &  0.46 &  $<$6.26 &      --- \\ 
1716        & 12 41 59.40 &  12 56 34.20 & \object{               NGC 4620} &  1125$\pm$3  &  0.99 & -18.06 &   --- &  9.22 &    --- &  0.46 &  $<$6.16 &      --- \\ 
1727        & 12 47 20.60 &  12 09 59.10 & \object{                IC 3779} &  1165$\pm$4  &  1.07 & -16.03 &   --- &  8.51 &    --- &  0.54 &  $<$6.27 &      --- \\ 
1766        & 12 58 09.00 &  14 51 32.10 & \object{              UGC 08081} &   853$\pm$16 &  0.83 & -14.90 &   --- &  7.94 &    --- &  0.83 &  $<$6.17 &      --- \\ 
1989        & 13 57 29.52 &  09 57 03.20 & \object{           CGCG 074-017} &  6969$\pm$2  &  1.30 & -19.82 & 10.29 &  9.98 & -11.92 &  0.40 &  $<$7.73 & $<$-2.56 \\ 
2028        & 14 04 45.70 &  14 22 55.30 & \object{               NGC 5454} &  7603$\pm$1  &  1.46 & -21.54 & 11.03 & 10.76 & -12.22 &  0.83 &  $<$8.12 & $<$-2.91 \\ 
2308        & 15 23 35.20 &  09 20 46.10 & \object{           CGCG 077-110} & 10703$\pm$2  &  1.48 & -21.62 & 11.22 & 10.77 & -12.38 &  0.75 &  $<$8.33 & $<$-2.89 \\ 
2345        & 15 57 08.10 &  22 24 16.40 & \object{               NGC 6020} &  4304$\pm$2  &  1.42 & -20.83 & 10.84 & 10.40 & -12.83 &  0.54 &  $<$7.43 & $<$-3.40 \\ 
2351        & 16 01 51.40 &  17 57 26.90 & \object{               NGC 6030} &  4403$\pm$1  &  1.47 & -20.76 & 10.74 & 10.39 & -12.03 &  0.54 &  $<$7.41 & $<$-3.34 \\ 
2355        & 16 02 19.80 &  16 20 45.70 & \object{              UGC 10144} & 11494$\pm$3  &  1.67 & -22.19 & 11.53 & 11.02 & -12.58 &  0.40 &  $<$8.16 & $<$-3.37 \\ 
2406        & 21 16 24.80 &  10 16 24.10 & \object{           CGCG 426-029} &  5175$\pm$2  &  1.26 & -19.54 & 10.15 &  9.87 & -11.09 &  0.45 &  $<$7.50 & $<$-2.65 \\ 
2418$^M$    & 21 31 37.60 &  11 49 53.90 & \object{           CGCG 426-062} &  8643$\pm$3  &  1.70 & -21.05 & 11.15 & 10.52 & -12.19 &  0.45 &  $<$7.91 & $<$-3.23 \\ 
2430        & 21 50 27.60 &  12 38 10.30 & \object{2MASX J21502753+1238103} &  6507$\pm$2  &  1.42 & -18.75 &  9.98 &  9.55 & -11.69 &  0.45 &  $<$7.67 & $<$-2.30 \\ 
2505        & 23 12 22.70 &  14 30 22.60 & \object{2MASX J23122267+1430223} &  4982$\pm$6  &  1.16 & -17.55 &  9.24 &  8.93 & -11.11 &  0.51 &  $<$7.54 & $<$-1.70 \\ 
2529        & 23 23 27.00 &  14 19 33.00 & \object{           CGCG 431-053} &  7651$\pm$2  &  1.60 & -20.54 & 10.86 & 10.34 & -12.27 &  0.60 &  $<$7.97 & $<$-2.89 \\ 
2530        & 23 24 23.50 &  15 26 36.30 & \object{2MASX J23242356+1526362} &  7658$\pm$2  &  1.49 & -19.70 & 10.41 &  9.96 & -11.86 &  0.60 &  $<$7.94 & $<$-2.47 \\ 
2535        & 23 25 33.60 &  14 15 15.00 & \object{             ASK 144275} &  3783$\pm$5  &  0.88 & -15.72 &  8.39 &  8.36 & -10.55 &  0.60 &  $<$7.36 & $<$-1.03 \\ 
2584        & 23 52 36.40 &  14 33 05.20 & \object{              UGC 12822} &  7939$\pm$2  &  1.50 & -21.37 & 11.04 & 10.62 & -12.03 &  0.71 &  $<$8.07 & $<$-2.97 \\ 
\label{tab:A3060_ND}
\end{longtable}
\tablefoot{Flags in Col. 1: $C$ for targets clearly confused with another galaxy within the Arecibo telescope beam, $M$ for marginal \nan\ detections, and $U$ for unreliable photometry.}
\end{landscape}
}

\end{document}